\documentclass[a4paper,twoside,12pt]{article}

\usepackage{graphicx,amstext,amsmath,amsbsy,amscd}
\usepackage[all]{xy}
\usepackage{latexsym}
\usepackage{cite}
\usepackage{graphicx}

\newcommand{\bplus}{^{(+)}}
\newcommand{\bminus}{^{(-)}}

\newcommand{\gthree}{{\gamma^3}}

\newcommand{\medsp}{\\[0.7ex]}

\newcommand{\tve}{\underline{\varepsilon}}

\newcommand{\ve}{\varepsilon}

\newcommand{\Ltext}[1]{\ensuremath{\itindex{\mathcal{L}}{#1}}}

\newcommand{\diff}[1][]{\mbox{d}#1}

\newcommand{\half}[1]{\ensuremath{\frac{#1}{2}}}
\newcommand{\intd}[1]{\int \!\! #1 \;}
\newcommand{\inv}[1]{\ensuremath{\frac{1}{#1}}}

\newcommand{\Stext}[1]{\itindex{\mathcal{S}}{#1}}

\newcommand{\derfrac}[2][]{\frac{\partial #1}{\partial #2}}

\newcommand{\itindex}[2]{\ensuremath{#1_{\mbox{\scriptsize{\itshape #2}}}}}

\DeclareMathOperator{\extdm}{d}
\newcommand{\extd}{\extdm \!}

\begin{document}
\renewcommand{\thefootnote}{\fnsymbol{footnote}}
\thispagestyle{empty}
\begin{titlepage}

\begin{flushright}
  TUW-03-21\\
  hep-th/0306217
\end{flushright}
\vspace{1cm}

{\centering \textbf{\large THE COMPLETE SOLUTION OF\\ 2D SUPERFIELD
    SUPERGRAVITY\\ FROM GRADED POISSON-SIGMA MODELS
\\ AND THE SUPER POINTPARTICLE}\large \par}
\vspace{0.5cm}

\begin{center}
 L. Bergamin\footnote{bergamin@tph.tuwien.ac.at} and W. Kummer\footnote{wkummer@tph.tuwien.ac.at\par}
\end{center}

{\centering \textit{Institute for Theoretical Physics, Vienna University
of Technology}\par}

{\centering \textit{Wiedner Hauptstra{\ss}e 8-10, A-1040 Vienna, Austria}\par}
\vspace{0.5cm}
\begin{abstract}
Recently an alternative description of 2d supergravities in terms
of graded Poisson-Sigma models (gPSM) has been given. As pointed out
previously by the present authors a certain subset of gPSMs can be
interpreted as {}``genuine'' supergravity, fulfilling the well-known
limits of supergravity, albeit deformed by the dilaton field. In our
present paper we show that precisely that class of gPSMs corresponds one-to-one
to the known dilaton supergravity superfield theories
presented a long time ago by Park and Strominger. Therefore, the unique
advantages of the gPSM approach can be exploited for the latter: We
are able to provide the first complete classical solution for any
such theory. On the other hand, the straightforward superfield formulation
of the point particle in a supergravity background can be translated
back into the gPSM frame, where {}``supergeodesics'' can be discussed
in terms of a minimal set of supergravity field degrees of freedom.
Further possible applications like the (almost) trivial quantization
are mentioned. 
\end{abstract}

\end{titlepage}


\renewcommand{\thefootnote}{\arabic{footnote}}
\setcounter{footnote}{0}
\numberwithin{equation}{section}

\section{Introduction}
Theories of gravity in 1+1 dimensions naturally emerge from the generalization
of (e.g. spherically) reduced Einstein gravity in arbitrary dimensions.
Decisive progress in the treatment of their classical and quantum
properties were the consequence of the discovery in the early 90-s
that a Cartan formulation in a specific light-like gauge \cite{Kummer:1992rt},
which is equivalent to an Eddington-Finkelstein gauge for the metric,
not only simplifies enormously the evaluation of the classical theory,
but even allowed an exact (trivial) nonperturbative quantization
\cite{Haider:1994cw,Schaller:1994pm,Schaller:1994np,Louis-Martinez:1994eh}.
After the application to a particular model with curvature and torsion
\cite{Katanaev:1990qm} it was realized that not only all 2d gravity models but an even larger class of theories
may be covered by the concept of Poisson-Sigma models (PSMs)
\cite{Schaller:1994uj,Schaller:1994es,Klosch:1996fi,Klosch:1996qv,Klosch:1996bw}.
There a set of target space coordinates (auxiliary fields on the 2d
world sheet) exists besides the gauge-degrees of freedom. In this framework
the simplicity of 2d classical and quantum gravity becomes manifest.
PSM models generalized naturally to the graded case (gPSM) when they are supplemented
by anti-commuting fields \cite{Strobl:1999zz,Ertl:2000si}. The resulting
models exhibit the typical gauge transformation of supergravity theories.
However, the fermionic extensions are highly ambiguous. In addition
they may introduce new singularities and/or obstructions as compared
to the bosonic theory for which they have been derived. This result
was obtained for the $N=(1,1)$ superextension, but should hold also for
higher $N$.

Recently the present authors realized \cite{Bergamin:2002ju} that a subset
of those gPSMs can be identified, which fulfills a constraint algebra
whose structure is very close to the algebra of {}``genuine'' supergravity.
In this algebra the modifications by the presence of the dilaton field
(and its single fermionic partner for $N=(1,1)$) are, in a sense, minimal.
The only gPSMs allowed by that algebra correspond to a \emph{unique}
class of (dilaton deformed) $N=(1,1)$ supergravity theories (called {}``minimal
field supergravity'', MFS\footnote{Already at this point the authors
  apologize for the introduction of quite a number of special acronyms. It
  seems that only in terms of those a  reasonably compact formulation of the
  strategy is possible (cf.\ also fig. \ref{fig:relations} below).}, in the following) in which -- somewhat
miraculously -- even all singularities and obstructions in the generic
fermionic extensions disappear. The bosonic part of physically interesting
theories (spherically reduced gravity \cite{Thomi:1984na,Hajicek:1984mz,Schmidt:1997mq,Schmidt:1998ih}, string inspired
black hole \cite{Callan:1992rs}, simplified models
\cite{Barbashov:1979bm,Teitelboim:1983ux,teitelboim:1984,jackiw:1984,Jackiw:1985je},
bosonic potential of supergravity from superspace \cite{Howe:1979ia}) are
special cases thereof. This is also a non-trivial result,
because the {}``potential'' of those bosonic theories must be derivable
from a prepotential.

Already in the purely bosonic case, where the PSM is equivalent \cite{Katanaev:1996bh,Katanaev:1997ni}
to a general 2d dilaton theory (GDT) with vanishing torsion but dynamical dilaton, the corresponding
PSM works with non-vanishing bosonic torsion. If this PSM action shall be
extended directly to its supersymmetrized version using the superspace formalism, the
generalization of the usual conventional constraints, valid solely for
vanishing bosonic torsion, is an imperative step. In consequence, a new solution of Bianchi identities
etc., has to be considered, which turned out to be a highly non-trivial task
\cite{Ertl:2001sj}. Within the
gPSM approach this problem is avoided altogether and it suffices to
solve a graded Jacobi-type identity (vanishing Nijenhuis tensor) \cite{Ertl:2000si}.

Therefore the question arises whether, and in what sense, the equivalence of
the bosonic PSM and GDT theories can be extended to supergravity. To this end
it must be investigated, whether a gPSM based MFS has any relation to a genuine dilaton superfield theory, expressed in terms
of superspace coordinates for a \emph{dynamical} super-dilaton field.
An indication that this may work comes from the known result
\cite{Ikeda:1993aj,Ikeda:1994fh,Izquierdo:1998hg,Ertl:2000si,Ertl:2001sj} that
a gPSM model with \emph{vanishing} bosonic torsion \cite{Izquierdo:1998hg} is --up to
elimination of auxiliary fields-- equivalent to a dilaton superfield theory \cite{Park:1993sd}
with \emph{non-dynamical} dilaton. By further elimination of an auxiliary spinor
(``dilatino'') this simpler model can be related quite generally to the supergravity
model of Howe \cite{Howe:1979ia} as well.

One of the main motivations to establish such a relation in the general case is the fact that
in the (g)PSM approach the complete exact classical solution can be
found for all such models. Also, for bosonic PSMs  the quantization is (almost) trivial
\cite{Kummer:1992rt,Haider:1994cw,Schaller:1994pm,Schaller:1994np}\footnote{For
  a comprehensive review we suggest \cite{Grumiller:2002nm}.} as long as no
matter interactions are included. But even with matter
a meaningful quantum perturbation theory can be developed
\cite{Kummer:1997hy,Kummer:1998jj,Kummer:1998zs}, leading
to an improved understanding of phenomena like the virtual black
hole \cite{Grumiller:2000ah,Grumiller:2002dm}. Most of these results should
extend straightforwardly to the gPSM \cite{Bergamin:2001}, which would allow
substantial progress in the understanding of generalized supergravity in two
dimensions.

In our present paper we are able to report that, indeed, a detailed
equivalence exists between
the class of gPSM-supergravities of reference \cite{Bergamin:2002ju} (MFS
models) and the well-known dilaton superfield supergravities,
proposed sometime ago by Park and Strominger \cite{Park:1993sd} (dubbed
{}``superfield dilaton supergravities'', SFDS). The equivalence
proceeds through different steps which should be transparent in the
schematic representation of fig.\ \ref{fig:relations}, an explanation thereof is given first:
\begin{itemize}
\item The two left hand columns of the figure cover the purely bosonic
  theories, the ones on the
  r.h.s.\ include their fermionic extensions. The two columns in the middle contain
  theories with dynamical dilaton, while
  the two columns at the borders are reserved for the restricted class of models with non-dynamical
  dilaton, respectively.
\item As indicated by arrows at the top of
  the figure, different theories displayed in a row are related to each other by means
  of supersymmetric extension or by restriction to non-dynamical dilaton.

  Two fermionic extensions (MFDS and SFDS) correspond to GDT, which is
  indicated by the large bracket.
\item Relations between different models are described by
  arrows. Double-headed arrows are used if the corresponding relation
  indicates complete equivalence, or, at least, holds for the most important class of the connected
  theories. Simple arrows point from the more general theories towards the
  restricted ones.
\item Labels with a tilde indicate that this relation is a straightforward
  generalization of the corresponding relation among bosonic theories (e.g.\
  $A \leftrightarrow \tilde{A}$). Relations among different arrows within the
  same part of the figure (bosonic part or supersymmetric part resp.)\ are
  indicated by primes (e.g.\ $A \leftrightarrow A'$).
\end{itemize}
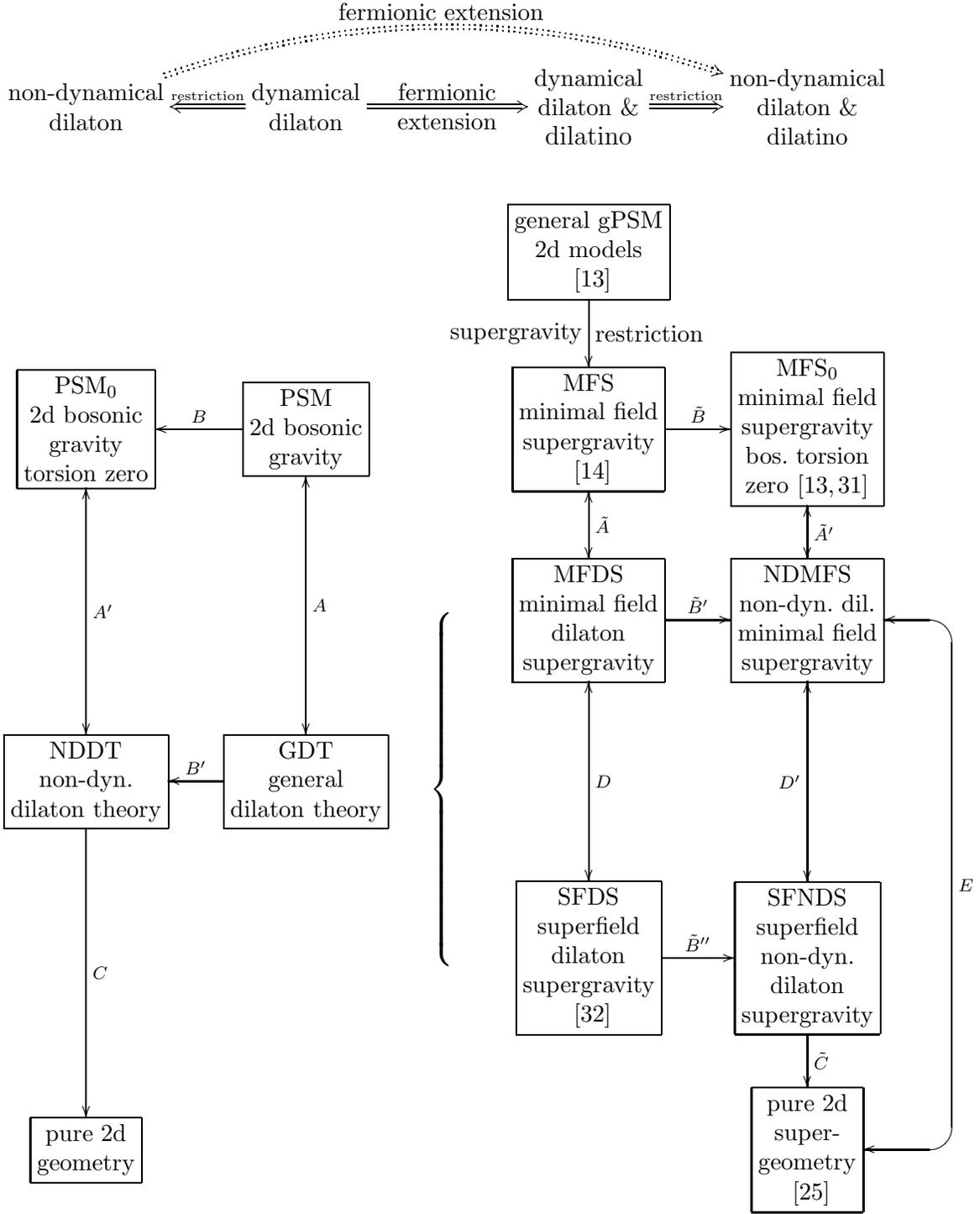
\begin{figure}[p]
\vspace{-20pt}
\label{fig:relations}
\[
\xymatrix{ \txt{\small non-dynamical\\\small dilaton} \ar @/^3pc/ @2{.>} [rrrr] ^-{\txt{\small
      fermionic extension}}  & \txt{\small
      dynamical\\\small dilaton} \ar @2{->} [l] _-{\txt{\tiny
      restriction}} \ar @2{->} [rr]^-{\txt{\small
      fermionic}}_-{\txt{\small extension}} & & \txt{\small dynamical\\\small dilaton \&\\
      dilatino} \ar @2{->} [r] ^-{\txt{\tiny
      restriction}} & \txt{\small
      non-dynamical\\\small dilaton \&\\\small dilatino} & \\
  &&& *+[F]\txt{\small general gPSM\\\small
   2d models\\\small \cite{Ertl:2000si}}  \ar[d]_-{\txt{\small supergravity}}^-{\txt{\small
      restriction}} & & \\
 *+[F]\txt{\small PSM$_0$\\\small 2d bosonic\\\small gravity\\\small torsion zero} \ar @{<-} [r] ^-B
      \ar @{<->} [dd] ^-{A'} & *+[F]\txt{\small
      PSM\\\small 2d bosonic\\\small gravity} \ar @{<->} [dd] ^-A & & *+[F]\txt{\small
      MFS\\\small minimal field\\\small supergravity\\\small \cite{Bergamin:2002ju} } \ar[r]
      ^-{\tilde{B}} \ar @{<->} [d] ^-{\tilde{A}} & *+[F]\txt{\small
      MFS$_0$\\\small minimal field\\\small supergravity\\\small bos.\
      torsion\\\small zero
      \cite{Izquierdo:1998hg,Ertl:2000si}} \ar @{<->} [d] ^-{\tilde{A}'}& \\
 & & & *+[F]\txt{\small
      MFDS\\\small minimal field\\\small dilaton\\\small supergravity} \ar @{<->} [dd] ^-D \ar[r] ^-{\tilde{B}'} & *+[F]\txt{\small
      NDMFS\\\small non-dyn.\ dil.\\\small minimal field\\\small supergravity}
    \ar @{<->} [dd] _-{D'}
    \ar @{<->} `[dr] `[ddd]^-E [ddd] & \\
 *+[F]\txt{\small NDDT\\\small non-dyn.\ \\\small dilaton theory} \ar[dd] ^-C \ar @{<-}
      [r] ^-{B'} & *+[F]\txt{\small
      GDT\\\small general\\\small dilaton theory} & & & & \\
 & & & *+[F]\txt{\small SFDS\\\small superfield\\\small dilaton\\\small supergravity\\\small
      \cite{Park:1993sd}} \ar[r] ^-{\tilde{B}''} & *+[F]\txt{\small
      SFNDS\\\small superfield\\\small non-dyn.\ \\\small dilaton\\\small supergravity} \ar  [d] ^-{\tilde{C}} & \\
*+[F]\txt{\small pure 2d\\\small geometry} & &  & & *+[F]\txt{\small pure
      2d\\\small super-\\\small geometry\\\small \cite{Howe:1979ia}}
\save "4,3"."6,3"!C * \frm{\{} \restore
  }
\]
\caption{Relation between different formulations of 2d gravity and
2d supergravity. Explanations are given in the text.}
\end{figure}

The equivalence $A$ between PSM and GDT (resp.\ $A'$ for non-dynamical dilaton) in the bosonic case is well-known
\cite{Katanaev:1996bh,Katanaev:1997ni,Grumiller:2002nm} so that $\tilde{A}$
amounts to a trivial generalization of $A$, when anti-commuting fields are
included \cite{Ertl:2000si}. It connects MFS to minimal field
dilaton supergravity (MFDS), the fermionic extension of GDT (the same is true
for $\tilde{A}'$).

The proof of the quite non-trivial equivalence $D$ provides the basis of our
present paper. We first establish it between the NDMFS and the SFNDS
theory, both \emph{without} dynamical dilaton, following the path $D'$  in fig. 1. In a second step SFNDS and SFDS are found to be connected by a (super-)conformal
transformation ($\tilde{B}''$ backwards), which in the gPSM frame possesses a
counterpart in a special target
space diffeomorphism (path $\tilde{B}$ backwards) between MFS$_0$ and MFS. That latter transformation
turns out to be a generalization of the conformal transformation
linking GDT and NDDT in the pure bosonic case (path $B$). We have found that in this way the more complicated direct relation of the general models (path $D$) is sufficiently transparent.
This strategy is especially important also for keeping track of the proper way the
symmetry transformations are mapped upon each other following those successive
steps.

Another equivalence is established between theories with non-dynamical dilaton
(MFS$_0$ and SFNDS resp.)\ and the model of Howe \cite{Howe:1979ia}. For the
restricted class of actions with invertible (pre-)potential this may be
obtained by the elimination of a superfield (path $\tilde{C}$) or,
alternatively, by the path $D' \rightarrow E$. This last equivalence also
allows to relate MFS$_0$ directly (i.e.\ without using $D'$, but instead
$\tilde{A}' \rightarrow E$) to Howe's supergravity \cite{Howe:1979ia}.

On the basis of those relations the technical advantages of gPSM
supergravity can be exploited in full detail for the SFDS theories
of ref. \cite{Park:1993sd}: Proceeding ``top down'' from the box MFS in
fig.\ \ref{fig:relations} ($\tilde{A} \rightarrow D$) and using the known general solution for the
MFS$_0$ \cite{Ertl:2000si} together with (the inverted arrow) $\tilde{B}$, we
are able to give the complete analytic solution
for the general superfield dilaton supergravity of ref. \cite{Park:1993sd},
including all fermionic contributions. 

Another example where the opposite way, the ``bottom up'' sequence ($D \rightarrow \tilde{A}$) is
to be chosen, is important for the determination of the supergravity
generalization of the geodesic within the gPSM formulation, because
the supersymmetric line element or the super-pointparticle can be defined straightforwardly in the superfield formulation
only.

The paper is organized as follows:
In Section \ref{sec:two} at first (Sect.\ \ref{sec:twoone}) the basic features of gPSMs are reviewed
shortly. Then (Sect.\ \ref{sec:twotwo}) the subset MFS of ``genuine'' supergravities is
described as determined in ref.\ \cite{Bergamin:2002ju} and the corresponding
MFDS (Sect.\ \ref{sec:twothree}) which obtains after elimination of certain
auxiliary\footnote{These ``auxiliary'' fields in the (g)PSM approach should
  not be confused with auxiliary fields in a superfield formulation.}
fields. They include the part of the spin-connection which, at the PSM level,
depends on the bosonic torsion and the target-space coordinates except the
dilaton and the dilatino.

Section \ref{sec:superfields} is devoted to the superfield approach of 2d supergravity, where it
suffices to consider the standard case with vanishing bosonic torsion. The
important r\^{o}le of (super-)conformal transformations is explained in Section \ref{sec:four}
which prepares the ground for the equivalence proof of minimal field
supergravity, as deduced from gPSMs, with dilaton superfield supergravity. The
proof is presented in
Section \ref{sec:torsiongrav}. All exact classical solutions of 2d superfield supergravity
\cite{Park:1993sd} are obtained in Section \ref{sec:solution}. Another application (Section
\ref{sec:sparticle}) is the formulation of a supergeodesic, defined as the
motion of a test particle in the background of minimal field
supergravity. Here only some very simple special cases are discussed, as e.g. the
null-directions  and the consequences for the supergravity background
generating the Schwarzschild solution. In
the Appendices we collect details of our notation and some lengthy formulas.

\section{Graded Poisson-Sigma Model and Minimal Field Supergravity}
\label{sec:two}
\subsection{Graded Poisson-Sigma Model}
\label{sec:twoone}
A general gPSM consists of scalar fields
$X^I(x)$, which are itself coordinates of a graded Poisson manifold with
Poisson tensor $P^{IJ}(X) = (-1)^{IJ+1} P^{JI}(X)$. The index
$I$, in the generic case, includes commuting as well as anti-commuting
fields\footnote{The usage of different indices as well as other features of
  our notation are explained in
  Appendix \ref{sec:notation}.
For further details one should consult ref.\ \cite{Ertl:2000si,Ertl:2001sj}.}. In addition one introduces the gauge
potential $A = \extd X^I A_I = \extd X^I A_{mI}(x) \extd x^m$, a one form with respect to the Poisson
structure as well as with respect to the 2d worldsheet coordinates. The gPSM
action reads\footnote{If the multiplication of forms is evident in what
  follows, the wedge symbol will be omitted.}
\begin{equation}
  \label{eq:gPSMaction}
  \begin{split}
    \Stext{gPSM} &= \int_M \extd X^I \wedge A_I + \half{1} P^{IJ} A_J \wedge
    A_I \medsp
    &= \int e \bigl(\partial_0 X^I A_{1 I} - \partial_1 X^I A_{0I} + P^{IJ}
    A_{0J} A_{1I} \bigr) \diff{^2 x}\ .
  \end{split}  
\end{equation}
The Poisson tensor $P^{IJ}$ must have vanishing Nijenhuis tensor (obey a
Jacobi-type identity with respect to the Schouten bracket related as $\{ X^I,
X^J \} = P^{IJ}$ to the Poisson tensor)
\begin{equation}
\label{eq:nijenhuis}
  P^{IL}\partial _{L}P^{JK}+ \mbox{\it 
g-perm}\left( IJK\right) = 0\ ,
\end{equation}
where the sum runs over the graded permutations. Due to
\eqref{eq:nijenhuis} the action \eqref{eq:gPSMaction} is invariant under the
symmetry transformations
\begin{align}
\label{eq:symtrans}
  \delta X^{I} &= P^{IJ} \ve _{J}\ , & \delta A_{I} &= -\mbox{d} \ve
  _{I}-\left( \partial _{I}P^{JK}\right) \ve _{K}\, A_{J}\ ,
\end{align}
where the term $\extd \epsilon_I$ in the second of these equations provides
the justification for calling $A_I$ ``gauge fields''.

For a generic (g)PSM the commutator of two transformations \eqref{eq:symtrans} is a
symmetry modulo the equations of motion (e.o.m.-s). Only for \( P^{IJ} \) linear
in \( X^{I} \) a closed (and linear) Lie algebra is obtained, and
\eqref{eq:nijenhuis} reduces to the Jacobi identity for the structure
constants of a Lie group. If the Poisson
tensor is singular --the actual situation in any application to 2d
(super-)gravity due to the odd dimension of the bosonic part of the tensor--
there exist (one or more) Casimir functions $C(X)$ obeying
\begin{equation}
\label{eq:casimir}
  \{ X^I, C \} = P^{IJ}\derfrac[C]{X^J} = 0\ ,
\end{equation}
which, when determined by the field equations of motion, are constants of
motion.
The variation of $A_I$ and $X^I$ in \eqref{eq:gPSMaction} yields the gPSM
field equations
\begin{align}
\label{eq:gPSeom1}
  \extd X^I + P^{IJ} A_J &= 0\ ,\medsp
\label{eq:gPSMeom2}
  \extd A_I + \half{1} (\partial_I P^{JK}) A_K A_J &= 0\ .
\end{align}
In the application to two dimensional $N = (1,1)$ supergravity\footnote{More complicated identifications of the 2d Cartan variables with
\( A_{I} \) are conceivable \cite{Strobl:2003}.}, the gauge
potentials comprise the spin connection $\omega_{ab} = \omega \epsilon_{ab}$, the zweibein and the gravitino:
\begin{align}
\label{eq:MFSvariables}
  A_I &= (A_\phi, A_a, A_\alpha) = (\omega, e_a, \psi_\alpha) & X^I &= (X^\phi,
  X^a, X^\alpha) = (\phi, X^a, \chi^\alpha)
\end{align}
The fermionic components \( \psi _{\alpha } \) (``gravitino'') and \( \chi
^{\alpha } \) (``dilatino'') are Majorana spinors. Local Lorentz invariance determines the $\phi$-components of the Poisson
tensor
\begin{align}
\label{eq:lorentzcov}
  P^{a \phi} &= X^b {\epsilon_b}^a\ , & P^{\alpha \phi} &= -\half{1}
  \chi^\beta {\gthree_\beta}^\alpha\ ,
\end{align}
and the supersymmetry transformation is encoded in $P^{\alpha \beta}$.
In a purely bosonic theory, the only arbitrary component of the
Poisson tensor is $P^{ab} = v \epsilon^{ab}$, where the locally Lorentz invariant 
``potential'' \( v=v\left( \phi ,Y\right)  \) describes different
models ( \( Y=X^{a}X_{a}/2 \) ). Evaluating \eqref{eq:gPSMaction} with that
$P^{ab}$ and $P^{a \phi}$ from \eqref{eq:lorentzcov} the action ($\epsilon =
\half{1} \epsilon^{ab} e_b \wedge e_a$ is the volume form, $De_a = \extd e_a +
\omega {\epsilon_a}^b e_b$)
\begin{equation}
  \label{eq:bosonicPSM}
  \Stext{PSM} = \int_{\mathcal{M}} \bigl( \phi \diff \omega + X^a D e_a +
  \epsilon v \bigr)
\end{equation}
is obtained. The physically most
interesting models are described by potentials quadratic in $X^a$
\begin{equation}
  \label{eq:bosonpot}
  v=Y\, Z\left( \phi \right) +V\left( \phi \right)\ .
\end{equation}
They include spherically reduced Einstein gravity
\cite{Thomi:1984na,Hajicek:1984mz,Schmidt:1997mq,Schmidt:1998ih}, the string
inspired
black hole \cite{Callan:1992rs}, the simplified model with \( Z=0 \) and
linear \( V\left( \phi \right)  \) \cite{Barbashov:1979bm,Teitelboim:1983ux,teitelboim:1984,jackiw:1984,Jackiw:1985je},
the bosonic part of the Howe model \cite{Howe:1979ia} etc.

Potentials of type \eqref{eq:bosonpot} allow the integration of the (single) Casimir function
$C$ in \eqref{eq:casimir}
\begin{align}
\label{eq:bosonicC}
  C&= e^{Q(\phi)} Y + W(\phi)\ , & Q(\phi) &= \int_{\phi_1}^\phi \extd \varphi
  Z(\varphi)\ , & W(\phi) &= \int_{\phi_0}^\phi \extd \varphi e^{Q(\varphi)}
  V(\varphi)\ ,
\end{align}
where e.g., in spherically reduced gravity $C$ on-shell is proportional
to the ADM-mass in the Schwarzschild solution.

The auxiliary variables
\( X^{a} \) and the torsion-dependent part of the spin connection $\omega$
can be eliminated by \emph{algebraic} equations of motion (path $A$ in fig.\ \ref{fig:relations}). Then the action
reduces to the familiar generalized dilaton theory
in terms of the dilaton field \( \phi  \) and the metric:
\begin{equation}
\label{eq:GDT}
  \Stext{GDT} = \intd{\diff{^2 x}} e \Bigl(\half{1} R \phi - \half{1} Z
  \partial^m \phi \partial_m \phi + V(\phi) \Bigr)
\end{equation}
Both formulations are equivalent at the classical \cite{Katanaev:1996bh,Katanaev:1997ni} as
well as at the quantum level
\cite{Kummer:1997hy,Kummer:1998jj,Kummer:1998zs}.

For theories with non-dynamical dilaton ($Z = 0$ in \eqref{eq:bosonpot}) a further
elimination of $\phi$ is possible if the potential $V(\phi)$ is
invertible. In this way one  arrives at a theory solely formulated in terms of
the zweibein $e_m^a$ (path $C$ in fig.\ \ref{fig:relations}).
\subsection{Minimal Field Supergravity}
\label{sec:twotwo}
For $N = (1,1)$ supergravity (cf.\ \eqref{eq:MFSvariables}) a generic
fermionic extension of the action \eqref{eq:bosonicPSM} is obtained 
by making general Lorentz invariant ans\"{a}tze for $P^{a \alpha}$, $P^{\alpha
  \beta}$ together with the
fermionic extension of $P^{ab} = \epsilon^{ab}(v + \chi^2 v_2)$ of the bosonic
case ($\chi^2 = \chi^\alpha \chi_\alpha$). Then the Jacobi identity 
\eqref{eq:nijenhuis} is solved. Here \eqref{eq:lorentzcov} and the bosonic potential $v$
are a given input. This leads to an algebraic, albeit highly
ambiguous solution with several arbitrary functions \cite{Ertl:2000si}. In addition, the
fermionic extensions generically exhibit new singular terms. Also
not all bosonic models permit such an extension for the whole range
of their bosonic fields, sometimes even no extension is allowed. 

As shown by the present authors \cite{Bergamin:2002ju}, it is, nevertheless,
possible to select {}``genuine'' supergravity from this huge set
of theories. This is possible by a generalization of the standard
requirements for a ``true'' supergravity
\cite{Freedman:1976xh,Freedman:1976py,Deser:1976eh,Deser:1976rb,Grimm:1978kp}
to the situation, where deformations from the dilaton field $\phi$ are present. To this end the
non-linear symmetry \eqref{eq:symtrans}, which is closed on-shell only, is
--in a first step-- related to a more convenient (off-shell closed) algebra of
Hamiltonian constraints $G^I = \partial_1 X^I + P^{IJ}(X) A_{1 I}$. The Hamiltonian obtained from
\eqref{eq:gPSMaction} in terms of these constraints takes the form
\cite{Grosse:1992vc,Haider:1994cw,Bergamin:2002ju}
\begin{equation}
H=\intd{\diff{x^{1}}}G^{I}A_{0I}\ .
\end{equation}
In a second step a certain linear combination of the \( G^{I} \), suggested by the
ADM parametrization \cite{Katanaev:1994qf,Katanaev:2000kc,Bergamin:2002ju}, maps the \( G^{I} \)
algebra upon a deformed version of the superconformal algebra (deformed
Neuveu-Schwarz, resp.\  Ramond algebra). This algebra is
appropriate to impose restrictions, which represent a natural
generalization of the requirements from supergravity to theories deformed by the
dilaton field. It turned out that the subset of
models allowed by these restrictions uniquely leads to the gPSM \emph{supergravity}
class of theories (called ``minimal field supergravity'', MFS, in our present paper) with the Poisson
tensor\footnote{The constant $\tilde{u}_0$ in ref.\ \cite{Bergamin:2002ju} has
  been fixed as
  $\tilde{u}_0 = -2$. This is in agreement with standard
  supersymmetry conventions.}
\begin{align}
\label{eq:mostgensup}
  P^{ab} &= \biggl( V + Y Z - \half{1} \chi^2 \Bigl( \frac{VZ + V'}{2u} +
  \frac{2 V^2}{u^3} \Bigr) \biggr) \epsilon^{ab}\ , \medsp
  P^{\alpha b} &= \frac{Z}{4} X^a
    {(\chi \gamma_a \gamma^b \gthree)}^\alpha + \frac{i V}{u}
  (\chi \gamma^b)^\alpha\ , \medsp
\label{eq:mostgensuplast}
  P^{\alpha \beta} &= -2 i X^c \gamma_c^{\alpha \beta} + \bigl( u +
  \frac{Z}{8} \chi^2 \bigr) \gthree^{\alpha \beta}\ ,
\end{align}
where the three functions $V$, $Z$ and the
``prepotential'' $u$ depend on the
dilaton field $\phi$ only. Besides the fixed components of $P^{IJ}$ according to
\eqref{eq:lorentzcov} supergravity requires the existence of supersymmetry
transformations, which are generated by the first
term in \eqref{eq:mostgensuplast}. It has been a central result of
ref.\ \cite{Bergamin:2002ju} that $P^{\alpha \beta}$ must be of the form
\eqref{eq:mostgensuplast}, i.e.\ the generator of supersymmetry
transformations
is not allowed to receive any deformations with respect to its form from
rigid supersymmetry. Furthermore in order to satisfy the condition
\eqref{eq:nijenhuis} $V$, $Z$ and $u$ must be related by ($u' = d
u/d \phi$)
\begin{equation}
\label{eq:finalpot}
V\left( \phi \right) =- \inv{8} \bigl(( u^{2})' + u^{2} Z\left( \phi
\right) \bigr) \ .
\end{equation}
Thus, starting from a certain bosonic model with potential \eqref{eq:bosonpot}
in \eqref{eq:mostgensup}, the only restriction remains that it must be expressible in
terms of a prepotential \( u \) by \eqref{eq:finalpot}. This happens to be the
case for most physically interesting theories
\cite{Thomi:1984na,Hajicek:1984mz,Schmidt:1997mq,Schmidt:1998ih,Callan:1992rs,Barbashov:1979bm,Teitelboim:1983ux,teitelboim:1984,jackiw:1984,Jackiw:1985je,Howe:1979ia}. Inserting the Poisson tensor \eqref{eq:lorentzcov},
\eqref{eq:mostgensup}-\eqref{eq:mostgensuplast}
into equation \eqref{eq:gPSMaction} the ensuing action becomes (the covariant
derivatives are defined in \eqref{eq:A8}) 
\begin{multline}
  \label{eq:mostgenaction}
  \Stext{MFS} = \int_{\mathcal{M}} \bigl( \phi \diff \omega + X^a D e_a + \chi^\alpha D
  \psi_\alpha + \epsilon \biggl( V + Y Z - \half{1} \chi^2 \Bigl( \frac{VZ + V'}{2u} +
  \frac{2 V^2}{u^3} \Bigr) \biggr) \medsp
   + \frac{Z}{4} X^a
    (\chi \gamma_a \gamma^b e_b \gthree \psi) + \frac{i V}{u}
  (\chi \gamma^a e_a \psi) \medsp
   + i X^a (\psi \gamma_a \psi) - \half{1} \bigl( u +
  \frac{Z}{8} \chi^2 \bigr) (\psi \gamma_3 \psi) \bigr)\ .
\end{multline}
For later reference we also define the simpler model with $\bar{Z} = 0$ (MFS$_0$),
where the fields are denoted by $(\bar{\phi}, \bar{X}^a, \bar{\chi}^\alpha)$
and  $(\bar{\omega}, \bar{e}_a, \bar{\psi}_\alpha)$
\begin{equation}
  \label{eq:MFS0action}
  \Stext{MFS$_0$}(\bar{\phi}, \bar{X}^a, \bar{\chi}^\alpha;
  \bar{\omega}, \bar{e}_a, \bar{\psi}_\alpha) = \bigl.\Stext{MFS}\bigr|_{Z=0;\,
  \mbox{\tiny fields} \rightarrow \overline{\mbox{\tiny fields}}}\ .
\end{equation}
In terms of \eqref{eq:lorentzcov} and
\eqref{eq:mostgensup}-\eqref{eq:mostgensuplast} the supersymmetry transformations of the MFS
model, according to \eqref{eq:symtrans} read:
\begin{align}
  \label{eq:gPSMtransf}
\delta \phi &= \half{1} (\chi \gthree \ve) \medsp
\delta X^a &= - \frac{Z}{4} X^b (\chi \gamma_b \gamma^a \gthree \ve) - \frac{i
  V}{u} (\chi \gamma^a \ve) \label{eq:gPSMtransf2}\medsp
\delta \chi^\alpha &= 2i X^c (\ve \gamma_c)^\alpha - \bigl(u + \frac{Z}{8}
\chi^2 \bigr) (\ve \gthree)^\alpha \label{eq:gPSMtransf3}\medsp
\delta \omega &= \frac{Z'}{4} X^b (\chi \gamma_b \gamma^a \gthree \ve) e_a + i
\bigl(\frac{V}{u}\bigr)' (\chi \gamma^a \ve) e_a + \bigl(u' + \frac{Z'}{8}
\chi^2 \bigr) (\ve \gthree \psi) \label{eq:gPSMtransf4}\medsp
\delta e_a &= \frac{Z}{4} (\chi \gamma_a \gamma^b \gthree \ve) e_b - 2i (\ve
\gamma_a \psi) \label{eq:gPSMtransf5}\medsp
\delta \psi_\alpha &= - (D \ve)_\alpha + \frac{Z}{4} X^a (\gamma_a \gamma^b
\gthree \ve)_\alpha e_b +  \frac{i
  V}{u} (\gamma^b \ve)_\alpha e_b + \frac{Z}{4} \chi_\alpha (\ve \gthree \psi)\label{eq:gPSMtransflast}
\end{align}
We list neither here nor below transformations with the three bosonic
parameters $\ve_i$. The symmetry transformation generated by
\eqref{eq:lorentzcov} corresponds to the local Lorentz transformations, the
other two, by the field-dependent choice of the symmetry parameter $\ve_a =
\xi^m A_{m a}$, describe 2d
diffeomorphisms $\xi^m$ \cite{Strobl:1999wv}. Clearly, the invariance with respect to
the latter three transformations is also evident from the explicit form of the
action \eqref{eq:mostgenaction}.
\subsection{Minimal Field Dilaton Supergravity}
\label{sec:twothree}
The PSM form of the action \eqref{eq:mostgenaction} represents a theory with non-vanishing
bosonic torsion. This can be seen easily from the e.o.m.\ obtained
by variation of \( X^{a} \). Nevertheless, it is (locally and globally)
equivalent to a theory with dynamical dilaton field and vanishing
bosonic torsion. We recall the basic steps of this relation (path $\tilde{A}$
in fig.\ \ref{fig:relations}) as applied already to
the gPSM in \cite{Ertl:2000si}.
For this purpose the
action \eqref{eq:mostgenaction} is most conveniently abbreviated as
\begin{equation}
\label{eq:gPSMshort}
  \Ltext{MFS} = \int_{\mathcal{M}} \bigl( \phi \diff \omega + X^a D e_a + \chi^\alpha D
  \psi_\alpha + \half{1} P^{AB}e_B e_A\bigr)\ ,
\end{equation}
where now $A = (a, \alpha)$ only includes the zweibein $e_a$ and the gravitino
$e_\alpha = \psi_\alpha$
components (cf.\ Appendix \ref{sec:notation}). Varying \eqref{eq:gPSMshort} with respect to $X^a$ leads to the
torsion equation
\begin{equation}
  D e_a + \half{1} (\partial_a P^{AB}) e_B e_A = 0\ ,
\end{equation}
which can be used to substitute the independent spin connection $\omega$ by the
dependent\footnote{Here and also in the superfield approach below supersymmetry covariant quantities acquire a
tilde, when they denote \emph{dependent} variables.} supersymmetry covariant connection $\tilde{\omega}$ and by the torsion
$\tilde{\tau}$:
\begin{align}
\label{eq:torsionelim}
  \omega_a &= e^m_a \omega_m =
  \tilde{\omega}_a - \tilde{\tau}_a \medsp
\label{eq:tildeomega}
  \tilde{\omega}_a &= \epsilon^{mn} \partial_n e_{ma} - i \epsilon^{mn} (\psi_n
  \gamma_a \psi_m) \medsp
\label{eq:tildetau}
  \tilde{\tau}_a &= - \half{1} (\partial_a \hat{P}^{AB})
  \epsilon^{mn} e_{Bn} e_{Am}\medsp
  \hat{P}^{AB} &= P^{AB} + 2i \delta^A_\alpha \delta^B_\beta X^c
  \gamma_c^{\alpha \beta}
\end{align}
By partially integrating the torsion dependent part of \eqref{eq:gPSMshort}
some derivatives are moved onto the dilaton field $\phi$ and the action reads (up to
total derivatives):
\begin{equation}
  \label{eq:gPSMshort2}
\begin{split}
  \Stext{MFS} &= \intd{\diff{^2 x}} e \biggl( \half{1} \tilde{R} \phi +
  (\chi \tilde{\sigma}) - \half{1} \hat{P}^{AB} \epsilon^{mn} e_{Bn}
  e_{Am}\medsp
  &\phantom{= \intd{\diff{^2 x}} e \biggl(}+ \bigl(X^a + e^a_m \epsilon^{mn} (\partial_n \phi) + \half{1}  e^a_m
  \epsilon^{mn} (\chi \gthree \psi_n) \bigr) \tilde{\tau}_a \biggr)
\end{split}
\end{equation}
The curvature scalar
\begin{align}
  \label{eq:Rtilde}
  \tilde{R}&=2 \ast \extd \tilde{\omega } = 2 \epsilon^{mn} \partial_n \tilde{\omega}_m
\end{align}
through
\eqref{eq:tildeomega} depends on the torsion free spin connection \( \tilde{\omega} \)
which, in turn, may be expressed as well by the metric $g_{mn} = e_m^a e_{n
  a}$. In addition, the fermionic partner of the curvature scalar has been
introduced, which is defined as
\begin{equation}
  \label{eq:susysigma}
  \tilde{\sigma}_\alpha = \ast (\tilde{D} \psi)_\alpha =
  \epsilon^{mn} \bigl( \partial_n \psi_{m\alpha} + \half{1} \tilde{\omega}_n (\gthree
  \psi_m)_\alpha \bigr)\ .
\end{equation}
Varying again with respect to $X^a$ finally allows to eliminate this field
as well:
\begin{equation}
\label{eq:genxa}
  X^a = - {e^a}_m \epsilon^{mn} \Bigl((\partial_n \phi) + \half{1} (\chi \gthree
  \psi_n) \Bigr)
\end{equation}
Inspecting the original
action \eqref{eq:gPSMshort} one realizes that this is the e.o.m.\  of the
independent spin connection $\omega$. It is important to notice
that the structure of \eqref{eq:genxa} does not depend on the details of the
Poisson tensor, but is determined solely by the condition of local Lorentz
invariance. Equations \eqref{eq:torsionelim} and \eqref{eq:genxa} are
algebraic and even linear in the variables to be eliminated. Therefore,
they may be reinserted into the action \eqref{eq:gPSMshort2}:
\begin{equation}
  \label{eq:gPSMshort3}
  \Stext{MFDS} = \intd{\diff{^2 x}} e \biggl(\half{1} \tilde{R} \phi +
  (\chi \tilde{\sigma}) - \half{1} \Bigl.\hat{P}^{AB}\Bigr|_{X^a} \epsilon^{mn} e_{Bn}
  e_{Am}\biggr)
\end{equation}
Here $X^a$ indicates that this field should be replaced by
\eqref{eq:genxa}.

Because in the MFS the Poisson tensor \( P^{ab} \) depends quadratically
on \( X^{a} \) (cf.\ \eqref{eq:mostgensup}), according to \eqref{eq:genxa} the usual quadratic
dynamical term for the dilaton field \( \phi  \) is produced. Thus,
reinserting the Poisson tensor \eqref{eq:mostgensup}-\eqref{eq:mostgensuplast} with
\eqref{eq:genxa} into \eqref{eq:gPSMshort3} yields the minimal field
\emph{dilaton} supergravity (MFDS):
\begin{multline}
\label{eq:mostgenaction2}
  \Stext{MFDS} = \intd{\diff{^2 x}} e \biggl( \half{1} \tilde{R} \phi +
  (\chi \tilde{\sigma}) + V - \inv{4 u} \chi^2 \bigl( V Z + V' + 4
  \frac{V^2}{u^2} \bigr) \medsp
  - \half{1} Z \Bigl( \partial^m \phi \partial_m \phi + \half{1} (\chi \gthree
  \psi^m) \partial_m \phi + \half{1} \epsilon^{mn} \partial_n \phi (\chi
  \psi_m)\Bigr) \medsp
  - \frac{i V}{u} \epsilon^{mn} (\chi \gamma_n \psi_m) + \half{u}
  \epsilon^{mn} (\psi_n \gthree \psi_m) \biggr)
\end{multline}
This action describes dilaton supergravity theories with minimal
field content and vanishing bosonic torsion: The bosonic variable
$e^a_m$ appears explicitly, but it also is contained in the dependent
spin connection \( \tilde{\omega } \) according to \eqref{eq:tildeomega}. Beside
the dilaton field \( \phi  \) the fermionic  dilatino \( \chi ^{\alpha } \)
remains. Clearly for $Z = 0$ (NDMFS in fig. \ref{fig:relations}) the dynamical terms for the
dilaton field disappear ($\bar{V} = \bar{V}(\bar{\phi})$
etc.):
\begin{multline}
\label{eq:simplifiedaction2}
  \Stext{NDMFS} = \intd{\diff{^2 x}} \bar{e} \biggl( \half{1} \Tilde{\Bar{R}} \bar{\phi} +
  (\bar{\chi} \Tilde{\Bar{\sigma}}) + \bar{V} - \inv{4 \bar{u}} \bar{\chi}^2 \bigl(\bar{V}' + 4
  \frac{\bar{V}^2}{\bar{u}^2} \bigr)  \medsp
  - \frac{i \bar{V}}{\bar{u}} \epsilon^{mn} (\bar{\chi} \gamma_n \bar{\psi}_m) + \half{\bar{u}}
  \epsilon^{mn} (\bar{\psi}_n \gthree \bar{\psi}_m) \biggr)
\end{multline}
While a further elimination of the dilatino is possible for quite
general NDMFS models (discussed in section
\ref{sec:torsionzeroequiv}), one can get rid of \( \phi  \)
in certain very special (simple) cases of \eqref{eq:simplifiedaction2} only,
namely for invertible
potential terms \( V, \) resp.\ \( u \) (cf.\ \eqref{eq:finalpot}).

The supersymmetry transformations of the MFDS model follow by eliminating $X^a$ and $\omega$ in
\eqref{eq:gPSMtransf}, \eqref{eq:gPSMtransf3}, \eqref{eq:gPSMtransf5} and
\eqref{eq:gPSMtransflast}. Except for \eqref{eq:gPSMtransflast} the new
transformation rules are immediate by substituting $X^a$ by
\eqref{eq:genxa}. In eq.\ \eqref{eq:gPSMtransflast} we use the explicit
formula of the covariant torsion (cf.\ \eqref{eq:tildetau} with
\eqref{eq:mostgensup}-\eqref{eq:mostgensuplast})
\begin{equation}
  \tilde{\tau}_a = - Z \bigl( X_a + \inv{4} (\chi \gamma_a \gamma^b \psi_n)
  e^n_b \bigr)\ .
\end{equation}
After some algebra the result
\begin{align}
  \delta \phi &= \half{1} (\chi \gthree \ve)\ , \label{eq:elimtransf}\medsp
  \delta \chi^\alpha &= - 2i \epsilon^{mn} \bigl( \partial_n \phi + \half{1}
  (\chi \gthree \psi_n) \bigr)  (\ve \gamma_m)^\alpha - \bigl(u + \frac{Z}{8}
\chi^2 \bigr) (\ve \gthree)^\alpha\ , \label{eq:elimtransf2}\medsp
  \delta {e_m}^a &= \frac{Z}{4} (\chi \gamma^a \gamma^b \gthree \ve) e_{mb} - 2i (\ve
\gamma^a \psi_m)\ , \label{eq:elimtransf3}\medsp
  \delta \psi_{m \alpha} &= - (\tilde{D} \ve)_\alpha + \frac{iV}{u} (\gamma_m
  \ve)_\alpha + \frac{Z}{4} \bigl( \partial^n \phi (\gamma_m \gamma_n
  \ve)_\alpha + \half{1} (\psi_m \gamma^n
  \chi) (\gamma_n \gthree \ve)_\alpha \bigr)
\label{eq:elimtransflast}
\end{align}
is obtained. 

The action \eqref{eq:mostgenaction2} with its symmetry transformations
\eqref{eq:elimtransf}-\eqref{eq:elimtransflast} is most
convenient for a comparison with a superfield formulation of 2d supergravity,
because in \eqref{eq:mostgenaction2} the bosonic torsion vanishes and it is precisely
this case for which the standard supergravity has been developed.

\section{Superfield Dilaton Supergravity}
\label{sec:superfields}
Any formulation of dilaton supergravity in superspace is embedded in the
background of pure 2d super-geometry.  The simplest non-trivial
superfield extension of the topological bosonic 2d action $\intd{\diff{^2 x}} e R$ is
obtained by promoting the determinant \( e = \sqrt{-g} \) to
the superdeterminant \( E \), and the curvature \( R \) to a component of a real
superfield \( S \), which appears in a function $\mathcal{F}(S)$. At the same
time the integration is extended to an integral over $N = (1,1)$ superspace
($z^M = (x^m, \theta^\mu)$):
\begin{equation}
  \label{eq:howe1}
  \Stext{Howe} = \intd{\diff{^2 x} \diff{^2 \theta}} E \mathcal{F}(S)
\end{equation}
In the following \eqref{eq:howe1} will be referred to as the {}``Howe-action'' because
the analysis of 2d supergravity in terms of superfields goes back to the
seminal paper \cite{Howe:1979ia} of this author.
In the notation and conventions of the Appendix (cf.\ also ref.\ \cite{Ertl:2000si}
and the superspace conventions of \cite{Ertl:2001sj}) the respective
\( \theta  \)-expansions read
%
\begin{align}
\label{eq:superdet}
  E &= e \bigl( 1 - 2i (\theta \underline{\zeta}) + \half{1} \theta^2 ( \underline{A} + 2 \underline{\zeta}^2 +
  \underline{\lambda}^2) \bigr)\ , \medsp
  S &= \underline{A} + 2 (\theta \gthree \tilde{\underline{\sigma}}) + 2 i \underline{A} (\theta \underline{\zeta}) + \half{1}
  \theta^2 \bigl( \epsilon^{mn} \partial_n \tilde{\underline{\omega}}_m - \underline{A} (\underline{A} + 2 \underline{\zeta}^2
  + \underline{\lambda}^2 ) - 4i (\underline{\zeta} \gthree \tilde{\underline{\sigma}}) \bigl)\ .
\end{align}
For reasons that will become clear in section \ref{sec:torsionequiv}
superfield components are consistently expressed by underlined letters, except
$e_m^a$. $\underline{\zeta}_\alpha$ and $\underline{\lambda}^a_\alpha$ are the components of the Lorentz
covariant decomposition of the gravitino $\underline{\psi}_a^\alpha = e^m_a
\underline{\psi}_m^\alpha$ according to eq.\
\eqref{eq:gravitinodecomp}. The dependent variables
$\tilde{\underline{\omega}}$, $\tilde{\underline{R}}$ and
$\tilde{\underline{\sigma}}$ are defined by the eqs.\ \eqref{eq:tildeomega},
\eqref{eq:Rtilde} and \eqref{eq:susysigma}, when substituting all variables
therein by underlined ones.

The independent variables in the Howe-action \eqref{eq:howe1} are the
components of the zweibein $e^a$, of its fermionic
partner $\underline{\psi}^a$ and an auxiliary field $\underline{A}$. Inserting
the decomposition \eqref{eq:gravitinodecomp} of $\underline{\psi}$ and integrating out superspace
eq.\ \eqref{eq:howe1} reduces
to (cf.\ \eqref{eq:thetaint}, derivatives with respect to $\underline{A}$ are indicated by a dot):
\begin{equation}
\label{eq:howe2}
\begin{split}
  \Stext{Howe} &= \intd{\diff{^2 x}} e \biggl( \half{1} \dot{\mathcal{F}}
  \tilde{\underline{R}} - \underline{A} (\underline{A} \dot{\mathcal{F}} - \mathcal{F}) + 2 \ddot{\mathcal{F}}
  \tilde{\underline{\sigma}}^2 - 2 i \underline{A} \ddot{\mathcal{F}} (\underline{\psi}^a \gamma_a
  \gthree \tilde{\underline{\sigma}})\medsp
  &\quad - \half{1} \underline{A}^2 \ddot{\mathcal{F}} (\underline{\psi}^m \underline{\psi}_m) +
  \bigl(\half{1} \underline{A}^2 \ddot{\mathcal{F}} - (\underline{A} \dot{\mathcal{F}} - \mathcal{F}) \bigr)
  \epsilon^{mn} (\underline{\psi}_n \gthree \underline{\psi}_m) \biggr)
\end{split}
\end{equation}
Here $\mathcal{F}(\underline{A})$ is $\bigl.\mathcal{F}(S)\bigr|_{\theta =
  0}$, the body of the function $\mathcal{F}(S)$ in
\eqref{eq:howe1}. The action \eqref{eq:howe2} remains invariant under the
supergravity transformations\footnote{In agreement with our systematic
  notation e.g.\ the covariant derivative $\underline{\tilde{D}}$ refers to the dependent
  spin connection \eqref{eq:tildeomega} for the underlined components of the superfield.} (as in the notation for the fields, $\tve$ is used to distinguish
  that transformation parameter from
$\ve$ in \eqref{eq:elimtransf}-\eqref{eq:elimtransflast}):
\begin{gather}
\label{eq:howetransf}
  \begin{alignat}{2}
    \delta {e_m}^a &= - 2i (\tve \gamma^a \underline{\psi}_m) &\qquad \delta {e^m}_a &= 2
    i (\tve \gamma^m \underline{\psi}_a)
  \end{alignat} \medsp
\label{eq:howetransf2}
    \delta {\underline{\psi}_m}^\alpha = - \bigl( (\tilde{\underline{D}} \tve)^\alpha + \half{i} \underline{A}
    (\tve \gamma_m)^\alpha\bigr) \medsp
\label{eq:howetransflast}
    \delta \underline{A} = -2 \bigl( (\tve \gthree \tilde{\underline{\sigma}}) - \half{i} \underline{A} {e^m}_a
    (\tve \gamma^a \underline{\psi}_m) \bigr)
\end{gather}
As it stands, \eqref{eq:howe2} cannot be equivalent to a more general supergravity
like $\Stext{MFDS}$ in \eqref{eq:mostgenaction2}. Only for
\eqref{eq:MFS0action}, the special case of a
non-dynamical dilaton, a relation will be worked out in section \ref{sec:torsionzeroequiv}, but
\eqref{eq:howe2} is clearly insufficient to represent the general theory with dynamical
dilaton field.

In order to describe the superfield generalization of all bosonic GDT with
\emph{dynamical} dilaton (as
exemplified by \eqref{eq:GDT}), $\phi$ is
promoted to a superfield as well and one arrives at the general superfield dilaton
supergravities (SFDS, cf.\ fig.\ \ref{fig:relations}) of Park and Strominger \cite{Park:1993sd}
\begin{equation}
  \label{eq:ps1}
  \Stext{SFDS} = \intd{\diff{^2x} \diff{^2 \theta}} E \bigl( J(\Phi) S +
  K(\Phi) D^\alpha \Phi D_\alpha \Phi + L(\Phi) \bigl)\ .
\end{equation}
The general dilaton supergravity model of this type is described by three
functions $J(\Phi)$, $K(\Phi)$ and $L(\Phi)$  of the dilaton superfield
\begin{equation}
  \label{eq:dilatonsf}
  \Phi = \underline{\phi} + \half{1} \theta \gthree \underline{\chi} + \half{1} \theta^2 \underline{F}\ .
\end{equation}
In \eqref{eq:dilatonsf} a scalar dilaton field \( \underline{\phi}  \) appears as the lowest
component. From superspace geometry the standard transformation rules \cite{Howe:1979ia,Ertl:2001sj}
\begin{align}
\label{eq:pstransf}
  \delta \underline{\phi} &= - \half{1} \tve \gthree \underline{\chi}\ , \medsp
\label{eq:pstransf2}
  \delta \underline{\chi}_\alpha &= - 2 (\gthree \tve)_\alpha \underline{F} + i (\gthree \gamma^b
  \tve)_\alpha (\underline{\psi}_b \gthree \underline{\chi}) - 2 i (\gthree \gamma^m \tve)_\alpha
  \partial_m \underline{\phi}\ , \medsp
\label{eq:pstransflast}
  \delta \underline{F} &= -2 i (\tve \underline{\zeta}) \underline{F} - \half{i} \bigl(\tve \gamma^m \gthree
  (\tilde{\underline{D}}_m \underline{\chi}) \bigr) + (\tve \underline{\lambda}^m) \bigl((\underline{\psi}_m \gthree \underline{\chi}) - 2
  \partial_m \underline{\phi} \bigr)
\end{align}
are an immediate consequence. Integrating out superspace and elimination
of the auxiliary fields \( \underline{F} \) and \( \underline{A} \) by their (algebraic)
e.o.m.-s is straightforward but leads to rather lengthy expressions.
We, therefore, relegate some relevant formulas to appendix \ref{sec:integrals}. Furthermore,
we assume in the following that the reparametrization \( J\left( \Phi \right) \rightarrow \Phi  \)
is possible, so that only \( K \) and \( L \) remain as two free
functions. This agrees with the appearance of only \( Z \) and \( V \) and with the simple
factor \( \phi  \) in front of \( \tilde{R} \)
in the bosonic part of $\Stext{MFDS}$ in \eqref{eq:mostgenaction2}\footnote{For models of the form
\eqref{eq:ps1} that do not allow a global reparametrization of this type, the
equivalence to a gPSM discussed below holds patch-wise, only.}.

Then \eqref{eq:ps1} becomes (\( L\left( \underline{\phi} \right)  \) and \( K(\underline{\phi})  \)
are the body of \( L(\Phi)  \) and \( K(\Phi)  \), derivatives
thereof are taken with respect to \( \underline{\phi} \))
\begin{equation}
  \label{eq:ps3}
  \begin{split}
    \Stext{SFDS} &= \intd{\diff{^2x}} e \biggl( \half{1} \tilde{\underline{R}} \underline{\phi} +
    (\underline{\chi} \tilde{\underline{\sigma}}) + 2 K \bigl(\partial^m \underline{\phi} \partial_m \underline{\phi} - \frac{i}{4} \underline{\chi} \gamma^m
  \partial_m \underline{\chi} - (\underline{\psi}_n \gamma^m \gamma^n \gthree \underline{\chi}) \partial_m \underline{\phi} \bigr) \medsp
    &\phantom{= \intd{\diff{^2x}} e \biggl(} + 2 K
    L^2 - L L' + L \epsilon^{mn} (\underline{\psi}_n \gthree \underline{\psi}_m)  + i L' (\underline{\zeta} \gthree \underline{\chi}) \medsp
    &\phantom{= \intd{\diff{^2x}} e \biggl(} + \inv{4} \bigl( \half{1} L'' - K' L  + K (\underline{\psi}_n \gamma^m \gamma^n
    \underline{\psi}_m) \bigr)
    \underline{\chi}^2 \biggl)\ ,
  \end{split}
\end{equation}
with the corresponding symmetry transformations
\begin{alignat}{2}
    \delta {e_m}^a &= - 2i (\tve \gamma^a \underline{\psi}_m)\ , &\qquad \delta {e^m}_a &= 2
    i (\tve \gamma^m \underline{\psi}_a)\ ,
\label{eq:psostransf}
\end{alignat}
\begin{align}
    \delta {\underline{\psi}_m}^\alpha &= - (\underline{\tilde{D}} \tve)^\alpha - \half{i} \bigl( 4 K
    L - L'  - \inv{4} K' \underline{\chi}^2\bigr)  
    (\tve \gamma_m)^\alpha\ , \medsp 
  \delta \underline{\phi} &= - \half{1} \tve \gthree \underline{\chi}\ , \medsp
\label{eq:psostransflast}
  \delta \underline{\chi}_\alpha &=  2 L (\gthree \tve)_\alpha  + i (\gthree \gamma^b
  \tve)_\alpha (\underline{\psi}_b \gthree \underline{\chi}) - 2 i (\gthree \gamma^m \tve)_\alpha
  \partial_m \underline{\phi}\ .  
\end{align}
We shall need below also the special case $K = 0$ of the action
\eqref{eq:ps3}, called SFNDS in fig.\ \ref{fig:relations}:
\begin{equation}
  \label{eq:ps4}
    \Stext{SFNDS} = \intd{\diff{^2x}} \Bar{e} \biggl( \half{1}
    \Tilde{\underline{\Bar{R}}} \underline{\Bar{\phi}} +
    (\underline{\Bar{\chi}} \Tilde{\underline{\Bar{\sigma}}}) - \Bar{L} \Bar{L}' + \Bar{L}
    \epsilon^{mn} (\underline{\Bar{\psi}}_n \gthree \underline{\Bar{\psi}}_m)  + i \Bar{L}'
    (\underline{\Bar{\psi}} \gthree \underline{\Bar{\chi}}) + \inv{8} \Bar{L}'' \underline{\Bar{\chi}}^2 \biggl)
\end{equation}
It is written in terms of barred variables variables $\underline{\Bar{\phi}}$, $\underline{\Bar{\chi}}$, $\Bar{e}^a_m$ and
  $\underline{\Bar{\psi}}$ in analogy to the notation of NDMFS, eq.\
  \eqref{eq:simplifiedaction2}.

The basic task (path $D$ in fig.\ \ref{fig:relations}) of Section \ref{sec:torsiongrav} is to show the equivalence of
$\Stext{MFDS}$ in \eqref{eq:mostgenaction2} with \( \Stext{SFDS} \) \eqref{eq:ps3}, \emph{together}
with a correct translation of the transformation laws
\eqref{eq:elimtransf}-\eqref{eq:elimtransflast} into \eqref{eq:psostransf}-\eqref{eq:psostransflast}. In view of the quite
different structures this clearly
has no obvious answer, although the number of fields and their type
(\( (e, \phi ,\psi,\chi) \) for MFDS, resp. \( (e, \underline{\phi}
,\underline{\psi}, \underline{\chi})  \)
for SFDS) coincide. Therefore, first the transformations connecting theories
``horizontally'' in fig.\ \ref{fig:relations} must be discussed.

\section{Target Space Diffeomorphisms and Conformal\\ Transformations}
\label{sec:four}
Transformations of fields in a certain action generically lead
to new theories when those transformations contain singularities.
A famous case is the string inspired black hole model \cite{Callan:1992rs}
which, even in interaction with minimally coupled matter, by a dilaton
field dependent (singular) conformal transformation can be brought to flat space.
In fact, this is the basic reason for being able to find the classical
solution in that model. The black hole singularity disappears in flat
space, and thus the global geometric properties of the theory experience
a profound change. Nevertheless, as long as such a transformation
is performed only locally in function space and if, at the end of the day, for
the physical interpretation one
returns to the variables of the original theory, this detour
can be a very valuable mathematical tool. 
\subsection{Target Space Diffeomorphism in gPSMs}
\label{sec:fourone}
Different gPSMs can be mapped upon each other by the target space
diffeomorphism
\begin{equation}
\label{eq:basicconf}
  X^I\ \  \Longrightarrow \ \ \bar{X}^I = \bar{X}^I(X)\ .
\end{equation}
It is straightforward to check that the action \eqref{eq:gPSMaction} is form-invariant under this
diffeomorphism when the gauge potentials and the Poisson tensor are transformed
according to
\begin{align}
\label{eq:gaugepotconf}
  \bar{A}_I &= \derfrac[X^J]{\bar{X}^I} A_J\ , \medsp
  \bar{P}^{IJ} &= \bigl( \bar{X}^I \stackrel{\leftarrow}{\partial}_K \bigr)
  P^{KL} \bigl( \stackrel{\rightarrow}{\partial}_L \bar{X}^J \bigr)\ .
\end{align}
Here $\stackrel{\rightarrow}{\partial}_I$ (cf.\ \eqref{eq:Ader}) is the usual derivative acting to
the right and $\stackrel{\leftarrow}{\partial}_I$ acts as
\begin{equation}
  f \stackrel{\leftarrow}{\partial}_I = (-1)^{I(f+1)}
  \stackrel{\rightarrow}{\partial}_I f\ .
\end{equation}
We emphasize again at this point that \eqref{eq:basicconf} need not hold globally
and thus physics may be different in two models connected by such
a transformation, when, e.g.\ in the case of gravity theories, the \( \bar{A}_I \) are identified
with the Cartan variables associated to the new gauge field coordinates.

The two models from $P^{IJ}$ and $\bar{P}^{IJ}$ clearly obey two \emph{different} sets of
symmetry transformations, cf.\ \eqref{eq:symtrans}. The relation among them can
be written as
\begin{align}
\label{eq:transftransf}
  \bar{\delta} \bar{X}^I &= \delta \bar{X}^I (X)\ , \medsp
  \bar{\delta} \bar{A}_I &= \delta \bar{A}_I (A, X) + \mbox{e.o.m.-s}
  \label{eq:transftransf2}\ , \medsp
  \bar{\ve}_I &= \derfrac[X^J]{\bar{X}^I} \ve_J \ .
\label{eq:transftransflast}
\end{align}
The necessity for the appearance of the e.o.m.-s in \eqref{eq:transftransf2} is easily
seen when inserting the transformed $\ve$ in the characteristic
derivative term of \eqref{eq:symtrans}:
\begin{equation}
  \delta \bar{A}_I (A, X) = - \derfrac[X^J]{\bar{X}^I} \extd \bigl(
  \derfrac[\bar{X}^K]{X^J} \bar{\ve}_K \bigr) + \ldots = - \extd \bar{\ve}_I - (-1)^K
  \derfrac[X^J]{\bar{X}^I} \extd \bigl(\derfrac[\bar{X}^K]{X^J}\bigr) \bar{\ve}_K + \ldots
\end{equation}
Obviously this produces terms of the form $\extd \bar{X}$, which are absent in
the rest of the transformation. This indicates that each $\extd \bar{X}^I$ has to be removed by
the e.o.m.-s \eqref{eq:gPSeom1} to arrive at the transformation law as given
in \eqref{eq:symtrans} for $\bar{\ve}(\ve,X)$. Finally we note that the e.o.m.-s \eqref{eq:gPSeom1} transform
into the same ones for $\bar{X}$ and $\bar{A}$, while the e.o.m.-s \eqref{eq:gPSMeom2}
transform into e.o.m.-s of both types, \eqref{eq:gPSeom1} and
\eqref{eq:gPSMeom2} in terms of $\bar{X}$ and $\bar{A}$.

It is worth mentioning a specialty of the gPSM structure at this point. In an action
based on linear symmetry transformations new related actions are usually obtained by a
rearrangement of invariant functions -- e.g.\ the rearrangement of superfields
to obtain the general Park-Strominger model from the special case with $K = 0$
as discussed below. On the other hand, the gPSM action is not
constructed by the composition of invariant functions and supergravity
invariant derivatives, but the invariant is always the whole
action. Thus, modifying a gPSM action \emph{necessarily} implies the
modification of the symmetry transformations (cf.\ \eqref{eq:transftransflast}).

\subsection{Conformal Transformation for MFS and MFDS}
\label{sec:gPSMcontrans}
In gPSM-theories conformal transformations are a special type of target space
diffeomorphisms. They, in particular, may be used
to connect (path $\tilde{B}$ in fig.\
\ref{fig:relations}) the MFS models \eqref{eq:mostgenaction} to MFS$_0$ models \eqref{eq:MFS0action} with
vanishing bosonic torsion ($\bar{Z} = 0$) or, equivalently, the MFDS models
\eqref{eq:mostgenaction2} to related models without dynamical dilaton (NDMFS, path $\tilde{B}'$ in fig.\
\ref{fig:relations}). The MFS$_0$ action \eqref{eq:MFS0action} is mapped upon
\eqref{eq:mostgenaction} of MFS by  (cf.\ ref.\ \cite{Ertl:2000si},
eqs.\ (5.42), (5.48))
\begin{align}
\label{eq:ct1}
  \phi &= \bar{\phi}\ , & X^a &= e^{- \half{1} Q(\phi) } \bar{X}^a\ , & \chi^\alpha
  &= e^{- \inv{4} Q(\phi)} \bar{\chi}^\alpha\ , \medsp
\label{eq:ct2}
  \omega &= \bar{\omega} + \half{Z} \bigl(\bar{X}^b \bar{e}_b + \half{1} \bar{\chi}^\beta \bar{\psi}_\beta
  \bigr)\ , & e_a &= e^{\half{1} Q(\phi)} \bar{e}_a\ , & \psi_\alpha &= e^{
  \inv{4} Q(\phi)} \bar{\psi}_\alpha\ ,
\end{align}
with $Q$ defined in \eqref{eq:bosonicC}.
After the fields $X^a$ and the part of $\omega$ dependent on bosonic torsion
have been eliminated the ensuing NDMFS action \eqref{eq:simplifiedaction2} is connected with the general MFDS action \eqref{eq:mostgenaction2} by
the same transformation rules for $\phi$, $\chi$ $e_a$ and $\psi$ as given in
\eqref{eq:ct1} and \eqref{eq:ct2}. The prepotential $u$ transforms according to
\begin{equation}
  \label{eq:prepottransf}
  u = e^{- \half{1} Q(\phi)} \bar{u}\ ,
\end{equation}
which leads to a canonical transformation of $\bar{V}(\phi) = - \inv{4}
\bar{u} \bar{u}'$ into \eqref{eq:finalpot}, such that the combination $\bar{e}
\bar{V} = e V$ remains invariant.

The symmetry transformations of the MFS models
\eqref{eq:gPSMtransf}-\eqref{eq:gPSMtransflast} with respect to the variables
with and without bar,
resp.,  are equivalent up to equations of motion of $\omega$ (or just as well
$\bar{\omega}$). In contrast, applying \eqref{eq:ct1}-\eqref{eq:ct2} to the symmetry transformations of the MFDS model
(\eqref{eq:elimtransf}-\eqref{eq:elimtransflast}) with $\bar{Z} = 0$
reproduces the the ones for $Z \neq 0$ without recourse to the e.o.m.-s of
$\omega$. Indeed, the latter have been used explicitely therein to eliminate the independent part of the spin connection.

In the NDMFS action \eqref{eq:simplifiedaction2} also the dilatino no longer
represents a dynamical field.  Variation with respect to $\bar{\chi}^\alpha$ leads to
\begin{equation}
  \label{eq:chielim}
  \bar{\chi}_\alpha = - \frac{8}{\bar{u}''} + 2i \frac{\bar{u}'}{\bar{u}''} {\epsilon_m}^n (\gamma^m
  \bar{\psi}_n)_\alpha\ ,
\end{equation}
and (provided $\bar{u}'' \neq 0$) the dilatino may be eliminated
altogether. The resulting action in terms of $e_a$, $\psi$ and $\phi$
($\Tilde{\Bar{R}}$ and $\Tilde{\Bar{\sigma}}$ are dependent variables as in
\eqref{eq:Rtilde}, \eqref{eq:susysigma}, but with $\bar{Z} = 0$) 
\begin{equation}
  \label{eq:dppaction2}
\begin{split}
  \Stext{NDMFS}^{(2)} &= \intd{\diff{^2 x}} \biggl( \half{1} \Tilde{\Bar{R}} \phi -
  \frac{4}{\bar{u}''} \Tilde{\Bar{\sigma}}^2 - \inv{8} (\bar{u}^2)' - 2i
  \frac{\bar{u}'}{\bar{u}''} \bar{e}^m_a (\bar{\psi}_m \gamma^a \gthree \Tilde{\Bar{\sigma}}) \medsp
  &\quad + \bigl( \half{\bar{u}} - \inv{4} \frac{(\bar{u}')^2}{\bar{u}''} \bigr) \epsilon^{mn}
  (\bar{\psi}_n \gthree \bar{\psi}_m) + \inv{4} \frac{(\bar{u}')^2}{\bar{u}''} (\bar{\psi}^m \bar{\psi}_m)
  \biggr)
\end{split}
\end{equation}
is invariant under the symmetry transformations
\begin{align}
\label{eq:dpptrans1}
  \delta {\bar{e}_m}^a &= - 2i (\ve \gamma^a \bar{\psi}_m)\ , \medsp
  \delta {\bar{\psi}_m}^\alpha &= - (\Tilde{\Bar{D}} \ve)^\alpha + \frac{i \bar{u}'}{4} (\ve
  \gamma_m)^\alpha\ , \medsp
  \delta \phi &= -\frac{4}{\bar{u}''} (\Tilde{\Bar{\sigma}} \gthree \ve) - \frac{i
  \bar{u}'}{\bar{u}''} \bar{e}^m_a (\bar{\psi}_m \gamma^a \gthree \ve)\ .
\label{eq:dpptranslast}
\end{align}
\subsection{Conformal Transformation in Superspace}
A similar conformal transformation connecting the general dilaton superfield
action SFDS \eqref{eq:ps1} to a model with non-dynamical
dilaton field (NDMFS, \( K=0 \) in \eqref{eq:ps1}) is also known in superspace
\cite{Park:1993sd} (path $\tilde{B}''$ in fig.\
\ref{fig:relations})\footnote{We re-emphasize that at the level of dilaton
  theories with vanishing bosonic torsion all known results
  \cite{Howe:1979ia,Park:1993sd} from 2d supergravity can be taken over.}.
It must contain a multiplication of the superzweibein $E_M^A$ with a factor
$\Lambda(\Phi)$ depending on the full
superspace multiplet \( \Phi  \). The resulting action is again an integral
over superspace. One consequence thereof is the fact that a kinetic term for
$\underline{\phi}$ necessarily implies a kinetic term for
$\underline{\chi}$. It is known that a super-Weyl
transformation preserving the constraints on the supertorsion (with vanishing
bosonic torsion) has the form
\cite{Howe:1979ia}
\begin{align}
  \label{eq:sspct}
  {E_M}^a &= \Lambda {\Bar{E}_M}^{\phantom{M}a}\ , & {E_M}^\alpha &= \Lambda^{\half{1}}
  {\Bar{E}_M}^{\phantom{M}\alpha} - i {\Bar{E}_M}^{\phantom{M}a} \gamma_a^{\alpha \beta} \underline{D}_\beta \Lambda^{\inv{2}}\ .
\end{align}
In our case we are
interested in the consequence of that transformation
on the action \eqref{eq:ps4} of SFNDS. Choosing in \eqref{eq:sspct}
  \begin{align}
    \label{eq:SPSWtrans}
    \Lambda &= \exp\bigl[\sigma(\Phi)\bigr]\ , & \sigma' &= -2 K\ ,
  \end{align}
in the action \eqref{eq:ps4} produces the general SFDS
action \eqref{eq:ps3}. The different components are related by
\begin{align}
  \label{eq:psct}
  \underline{\phi} &= \underline{\Bar{\phi}}\ , & \underline{\chi} &= e^{-\half{\sigma}} \underline{\Bar{\chi}}\ , &
  e_m^a &= e^\sigma \Bar{e}_m^a\ ,
\end{align}
\begin{equation}
  \label{eq:pspsict}
   \underline{\psi}_m^\alpha = e^{\half{\sigma}}
  \underline{\Bar{\psi}}_m^\alpha - \half{i} K e^{\half{\sigma}} \Bar{e}_m^a
  (\underline{\Bar{\chi}} \gamma_a \gthree)^\alpha\ ,
\end{equation}
where for $\sigma$ resp.\ $K$ the body $\sigma(\underline{\phi})$, resp.\ $K(\underline{\phi})$, is understood. 

\section{Equivalence of gPSM and Superfield Dilaton\\ Supergravity}
\label{sec:torsiongrav}
\subsection{Equivalence for Non-Dynamical Dilaton}
\label{sec:torsionzeroequiv}
Inspection of the SFNDS superfield action \eqref{eq:ps4} without dynamical dilaton
and of the action $\Stext{NDMFS}$ in \eqref{eq:simplifiedaction2}, which
originated from the gPSM formulation of supergravity, shows that in this special case the two theories
are the same, identifying
\begin{align}
\label{eq:dppequiv}
  \underline{\Bar{\psi}} &= \bar{\psi}\ , & \underline{\Bar{\chi}} &=
  \bar{\chi}\ , & \underline{\phi} &= \phi\ , & \Bar{L}(\phi) &=
  \half{\bar{u}(\phi)}\ .
\end{align}
It can be checked straightforwardly that the equivalence holds as well at the level of
the symmetry transformations when $\ve$ and $\tve$ are
identified (cf.\ eqs.\ \eqref{eq:elimtransf}-\eqref{eq:elimtransflast} with $Z
= 0$ and \eqref{eq:psostransf}-\eqref{eq:psostransflast} with $K=0$). Indeed, this relation of the Park-Strominger
model with $K = 0$ to a gPSM (interpreted as model with nonlinear
super-Poincar\'{e} algebra) had
been observed already before \cite{Ikeda:1994dr,Ikeda:1994fh}. In the
gPSM-based formalism the identification corresponds to the sequence of paths $\tilde{A}' \rightarrow
D'$ in fig.\ \ref{fig:relations}
\cite{Izquierdo:1998hg,Ertl:2000si}.

For a non-dynamical dilaton $\phi$ we observe yet another
identification between $\Stext{NDMFS}$ of eq.\
\eqref{eq:simplifiedaction2} and $\Stext{Howe}$ of \eqref{eq:howe2}, when
the dilatino in the former case has been eliminated as in eq.\
\eqref{eq:dppaction2}. Indeed eqs.\ \eqref{eq:dppaction2} and \eqref{eq:howe2}, as well
as the corresponding symmetry transformations
\eqref{eq:dpptrans1}-\eqref{eq:dpptranslast} and
\eqref{eq:howetransf}-\eqref{eq:howetransflast} are identical for
\begin{align}
\label{eq:gpsmhowe}
  \underline{\psi} &= \bar{\psi}\ , & \underline{A} &= - \frac{\bar{u}'}{2}\ , &
  \mathcal{F}(\underline{A}) &= \half{1} \Bigl(
  \bar{u}\bigl(\phi(\underline{A})\bigr) - \phi(\underline{A})
  \bar{u}'\bigl(\phi(\underline{A})\bigr)\Bigr)\ .
\end{align}
In equation \eqref{eq:chielim} we had to assume that $u'' \neq 0$ and thus the
invertibility of the first equation of \eqref{eq:gpsmhowe} is
guaranteed.

The equivalence \eqref{eq:gpsmhowe} corresponds to the
steps $\tilde{A}' \rightarrow E$ in figure \ref{fig:relations}. Alternatively
the path $E$ establishes a relation $D' \rightarrow E$ between two superspace actions, namely superfield dilaton supergravity with
non-dynamical dilaton (SFNDS, eq.\ \eqref{eq:ps4}) and the model of Howe
\eqref{eq:howe2}. $D' \rightarrow E$ and $\tilde{C}$ are not identical: Relation
$\tilde{C}$ can entirely be formulated in superspace and holds (as its bosonic
counterpart $C$, cf.\ comment below \eqref{eq:GDT}) in very special cases only (invertible potentials or
prepotentials, resp.). On the other hand, the path $D' \rightarrow E$
does not correspond to the elimination of a superfield. Instead the two
superfields in the version of \eqref{eq:ps1} with\footnote{This action
  corresponds to the sum of \eqref{eq:appexp1} and \eqref{eq:appexp3}.} $K = 0$, $\Bar{E}^a = (\Bar{e}^a, \underline{\Bar{\psi}}^a, \underline{\Bar{A}})$
and $\Bar{\Phi} = (\underline{\Bar{\phi}}, \underline{\Bar{\chi}}, \underline{\Bar{F}})$, are related to the
superfield in the model of Howe $E^a =  (e^a, \psi^a, A)$ by the steps
\[
\begin{array}{ccc}
  \mbox{SFNDS } \begin{matrix}
    (\Bar{e}^a, \underline{\Bar{\psi}}^a, \underline{\Bar{A}}) \\ (\underline{\Bar{\phi}}, \underline{\Bar{\chi}}, \underline{\Bar{F}})
  \end{matrix} & \xrightarrow[D']{\mbox{\tiny elimination of $\underline{\Bar{A}}$ and
    $\underline{\Bar{F}}$}} & 
  \mbox{NDMFS}\begin{matrix}
    (\Bar{e}^a, \underline{\Bar{\psi}}^a) \\ (\underline{\Bar{\phi}}, \underline{\Bar{\chi}})
  \end{matrix}\ \ , \\[4ex]
  \mbox{NDMFS }\begin{matrix}
    (\Bar{e}^a, \underline{\Bar{\psi}}^a) \\ (\underline{\Bar{\phi}}, \underline{\Bar{\chi}})
  \end{matrix} & \xrightarrow[E]{\mbox{\parbox{2.9cm}{\tiny  elimination of $\underline{\Bar{\chi}}$\\ reinterpretation $\underline{\Bar{\phi}}
    \rightarrow A$}}} &  \mbox{Howe } (e^a, \psi^a, A)\ \ .
\end{array}
\]
Obviously this equivalence combines components of different superfields in
\eqref{eq:ps1} into the components of the superfield $S$ of \eqref{eq:howe1}. Whenever the last
equation in \eqref{eq:gpsmhowe} can be solved explicitly for $A$, $D'
\rightarrow E$ is equivalent to $\tilde{C}$. However, $\tilde{C}$ is
meaningful if and only if such a solution exists, while eq.\
\eqref{eq:dppaction2} -- the result of $D' \rightarrow E$ -- does not depend on
the latter.  
\subsection{Dynamical Dilaton}
\label{sec:torsionequiv}
One may think that by means of (super-)conformal transformations, proceeding along the paths
$\tilde{B}$ resp.\ $\tilde{B}''$, also in the general case the identification
(path D)
can be established in a straightforward manner. However, with a
\emph{dynamical} dilaton the problem remains how to relate the
fields $(\psi,\chi)$, resp.\ $(\underline{\psi}, \underline{\chi})$, because no
obvious identification thereof
is apparent. Also the relation between the symmetry
transformations is far from trivial. Indeed, comparison of \eqref{eq:mostgenaction2} and
\eqref{eq:elimtransf}-\eqref{eq:elimtransflast} with \eqref{eq:ps3} and
\eqref{eq:psostransf}-\eqref{eq:psostransflast} immediately leads to the
two important observations:
\begin{itemize}
\item While the SFDS action \eqref{eq:ps3} includes standard kinetic terms for
   both, the dilaton field $\underline{\phi}$
  and its supersymmetric partner $\underline{\chi}$, in the MFDS formulation for $\phi$ such a
  term is generated too, but not for the dilatino.
\item The transformations of the zweibeine, \eqref{eq:psostransf} in the SFDS
  action and \eqref{eq:elimtransf3} in the MFDS action, are different. But in any comparison of the two models
  we had to assume that the zweibeine should be the same. Thus 
  the gravitini $\psi$ and $\underline{\psi}$ appearing on the r.h.s.\ of
  these transformations must be different.
\end{itemize}
In contrast to
the super-Weyl transformation in superspace (path $\tilde{B}''$ in fig.\
\ref{fig:relations}), the conformal transformation leading
from MFS$_0$ \eqref{eq:MFS0action} with $Z = 0$ to the MFS \eqref{eq:mostgenaction2} ($ Z
\neq 0$, path $\tilde{B}$ in fig.\
\ref{fig:relations}) depends on the dilaton field $\phi$ alone (cf.\ \eqref{eq:ct1}
and \eqref{eq:ct2}). One could, of course, try to
introduce more complicated transformations including also a dependence on the
dilatino $\chi$. It
turns out, that this is neither necessary nor possible: as pointed out
above, the relation \eqref{eq:genxa}, leading to the kinetic term for
$\phi$, does not depend on the details of the Poisson tensor and hence not on
a pecularity of some special class of models. In fact \emph{any} gPSM
with local Lorentz invariance after elimination of $X^a$ and
$\omega$ exhibits at best a kinetic term in $\phi$, but never in $\chi$. A similar conclusion holds
for the symmetry transformations (cf.\ the comment at the end of
section \ref{sec:fourone}).

On the other hand, the missing kinetic term for $\chi$ in MFDS could be generated by an appropriate
mixing of $\psi$ and $\chi$ in $\underline{\psi}$.  It is not difficult to find the correct
relation. When the SFNDS model \eqref{eq:ps4} is transformed according to
\eqref{eq:psct} one could try to replace \eqref{eq:pspsict} by the simpler rule
\begin{equation}
\label{eq:zetamod}
  \psi = e^{\half{\sigma}} \underline{\Bar{\psi}}\ .
\end{equation}
This implies a new definition of the gravitino $\psi$. On the other hand, in
this way a connection with the MFDS action \eqref{eq:mostgenaction2}, the one
following from the gPSM approach, can be established. Namely, identifying the
gravitino $\psi$ in \eqref{eq:zetamod} with the gravitino of that action
produces all terms there, provided
\begin{align}
\label{eq:genequiv}
  \underline{\phi} &= \phi\ , & \underline{\chi} &= \chi\ ,  & K(\phi) &= -
  \inv{4} Z (\phi)\ , & L(\phi) &= \half{u}\ .
\end{align}
Now all terms on the l.h.s\ of \eqref{eq:genequiv} refer to superfield
supergravity, whereas on the r.h.s.\ we find quantities defined in the
gPSM-based MFDS approach. This is not surprising, as the transformation rules \eqref{eq:psct}
together with \eqref{eq:zetamod} of SFDS by taking into account $K
= -\inv{4} Z$ are equivalent\footnote{Notice the similarity between
  \eqref{eq:bosonicC} and \eqref{eq:SPSWtrans}.} to the transformations of $\phi$, $e$,
$\chi$ and $\psi$ in \eqref{eq:ct1} and \eqref{eq:ct2}. 

So far we followed the paths $D' \rightarrow \tilde{B}'$ in fig.\
\ref{fig:relations}. In order to establish the relation $D$ between SFDS and MFDS,
the main goal of this section, the ansatz
\begin{equation}
  \label{eq:psi1}
  \underline{\psi}_m^\alpha = \psi_m^\alpha - \frac{i}{8} Z(\phi) e^a_m
  \epsilon_{ab} (\chi \gamma^b)^\alpha
\end{equation}
together with \eqref{eq:genequiv} suggests itself by comparison of
\eqref{eq:zetamod} with \eqref{eq:pspsict}. It follows when the conformal factors in the two terms on
the r.h.s.\  of \eqref{eq:pspsict} are absorbed first in a redefinition of
$\underline{\bar{\psi}}$; then the conformal transformations \eqref{eq:ct1} and \eqref{eq:ct2} for
$\chi^\alpha$, $e_a$ and $\psi_\alpha$ are taken into account. Not surprisingly, all contributions to
\eqref{eq:mostgenaction2} linear in $Z$ are reproduced. But also terms proportional $Z^2$ are
found to cancel.

There seems to remain a difference in the symmetry transformations. Assuming $\tve =
\ve$ one obtains ($\Delta = \itindex{\delta}{MFDS} - \itindex{\delta}{SFDS}$)
\begin{align}
\label{eq:Deltatrans}
  \Delta \phi &= 0\ , & \Delta \chi_\alpha &= \frac{Z}{8} \chi^2 (\gthree
  \ve)_\alpha\ , & \Delta e_m^a &= \half{Z} \epsilon^{ab} \chi \ve e_{mb}\ , &
  \Delta \psi_{m \alpha} = - \frac{Z}{4} \chi \ve (\gthree \psi_m)_\alpha\ .
\end{align}
However, this is nothing else but a local Lorentz transformation (cf.\ eq.\ \eqref{eq:A8}) with field dependent parameter
$\ve_{\phi} = \half{Z} \chi \ve$. An analogous transformation
emerges as well in the gPSM based formalism: the application of  \eqref{eq:transftransflast} leads
to ($\bar{\ve}$ denotes the symmetry
parameter of the MFS$_0$ model with $\bar{Z} = 0$) 
\begin{align}
    \ve_\phi &= \bar{\ve}_\phi + \half{Z} \bigl(X^b \ve_b + \half{1} \chi^\beta \ve_\beta
  \bigr)\ , & \ve_a &= e^{\half{1} Q(\phi)} \bar{\ve}_a\ , & \ve_\alpha &= e^{\inv{4}
  Q(\phi)} \bar{\ve}_\alpha\ .
\end{align}
The superspace parameters $\tve$ obey a similar relation, but with the opposite sign in
front of $Z/2$ in the first equation. Thus, a supersymmetry
transformation $\bar{\ve}_\alpha = \bar{\tve}_\alpha$ in MFS$_0$, resp.\
SFNDS, under the conformal transformations \eqref{eq:ct1}, \eqref{eq:ct2} and
\eqref{eq:psct} becomes ($\ve = (\ve_\phi, \ve_a, \ve_\alpha)$)
\begin{align}
  \bar{\ve} &= (0,0,\bar{\ve}_\alpha) &\longrightarrow&& \ve &= (\frac{Z}{4}
  \chi\ve,0 , \ve_\alpha )\ ,\medsp
  \bar{\tve} &= (0,0,\bar{\tve}_\alpha) &\longrightarrow&& \ve &= (- \frac{Z}{4}
  \chi\tve,0 , \tve_\alpha )\ .  
\end{align}

Adding the two contributions to $\ve_\phi$ and $\tve_\phi$ resp.\ yields the
result found in \eqref{eq:Deltatrans}: $\ve_{\phi}
= \half{Z} \chi \ve$. This terminates the proof that the minimal field supergravity in the
sense of ref.\ \cite{Bergamin:2002ju} is -- up to elimination of auxiliary fields --
equivalent to SFDS, the superfield dilaton gravity of Park and Strominger. The
symmetry transformations are mapped correctly upon each other, modulo a local
field-dependent Lorentz transformation.

It may be useful to conclude this section with a compilation of the relevant
formulas which, in agreement with the corresponding sequence of steps in fig.\
\ref{fig:relations}, relate minimal field  supergravity with the superfield
dilaton theory of ref.\ \cite{Park:1993sd}.
\begin{description}
\item[Actions:] Seven different actions that describe in some sense 2d
  supergravity have been presented. These are
  \begin{enumerate}
  \item the gPSM based MFS of eq.\ \eqref{eq:mostgenaction} and the special
    version MFS$_0$ thereof with vanishing bosonic torsion \eqref{eq:MFS0action},
  \item general dilaton supergravity MFDS in \eqref{eq:mostgenaction2}
    with its special version with non-dynamical dilaton
    \eqref{eq:simplifiedaction2} (NDMFS),
  \item SFDS of eq.\ \eqref{eq:ps3} and SFNDS of eq.\ \eqref{eq:ps4}, which
    both originate from the general dilaton superfield theory by Park and
    Strominger \eqref{eq:ps1},
  \item the model of Howe in eqs.\ \eqref{eq:howe1} and \eqref{eq:howe2}
    which, when derived from NDMFS \eqref{eq:simplifiedaction2} by elimination of the dilatino, takes the
    form \eqref{eq:dppaction2}.
  \end{enumerate}
\item[Transformations:] The dilaton field $\phi$ and the zweibeine $e^a_m$
  coincide for all models.
\begin{description}
\item[path $\mathbf{\tilde{A}}$:] The MFS fields ($\phi$, $X^a$, $\chi^\alpha$) and
  ($\omega$, $e_a$, $\psi_\alpha$) are reduced to the set ($\phi$, $\chi^\alpha$,
  $e_a$, $\psi_\alpha$) of MFDS by \eqref{eq:torsionelim}-\eqref{eq:tildetau}
  and \eqref{eq:Rtilde}-\eqref{eq:genxa}.
\item[paths $\mathbf{\tilde{B},\tilde{B}',\tilde{B}''}$:] At each level (MFS,
  MFDS and SFDS) a special target space transformation connects the models with
  non-dynamical dilaton (barred variables: MFS$_0$, MFNDS, SFNDS) to the general ones. For MFS
  this relation turns out to be the conformal transformation of eqs.\
  \eqref{eq:ct1} and \eqref{eq:ct2} which, when restricted to the fields $(\phi, \chi^\alpha,
  e_a, \psi_\alpha )$, also holds for MFDS vs.\ NDMFS. For SFDS the
  super-Weyl transformations \eqref{eq:sspct} and \eqref{eq:psct}, \eqref{eq:pspsict} are applied.
\item[path $\mathbf{D}$:] After elimination of the auxiliary fields in SFDS
  (eqs.\ \eqref{eq:psauxfield1} and \eqref{eq:psauxfield2}), this theory is
  equivalent to MFDS: the identification of the remaining fields and (pre-)potentials is
  contained in eqs.\ \eqref{eq:genequiv} and \eqref{eq:psi1}, the supersymmetry
  transformations are equivalent up to a local Lorentz transformation
  \eqref{eq:Deltatrans}.
\item[path $\mathbf{E}$:] The NDMFS action allows the elimination of the dilatino (eq.\
  \eqref{eq:chielim}), leading to a theory that may be identified with the
  model of Howe (eq.\ \eqref{eq:gpsmhowe}). Only in certain cases path $E$ is equivalent
  to the superfield relation $\tilde{C}$. Therefore, a combination of the
  paths $E \rightarrow \tilde{C}$ cannot be used as an alternative to $D'$.
\end{description}
\end{description}

\section{Solution of the General Dilaton Supergravity\\ Model}
\label{sec:solution}
The close relation between the general gPSM describing MFS supergravity and the general
superfield supergravity \eqref{eq:ps1} can be used to combine the advantages of both
approaches. We first use the fact that in MFS, as a gPSM, it is manifestly simpler
to arrive at the complete (classical and quantum)
solution of a 2d gravity system. Using the list of formulas described in the
last paragraph of the preceding section it can be mapped directly into the
complete exact solution of the Park-Strominger supergravity \eqref{eq:ps1},
where we assume that a redefinition of $\Phi$ by the replacement $J(\Phi)
\rightarrow \Phi$ is possible everywhere.

As supergravity in two dimensions without
matter has no propagating degrees of freedom
the physical content of the system is encoded in the Casimir functions \eqref{eq:casimir}. Every
gPSM gravity possesses at least one Casimir function, as the bosonic part of the
tensor has odd dimension (cf.\ \eqref{eq:bosonicC}). For the MFS$_0$ model (\eqref{eq:mostgenaction} with
$Z = 0$) this function can be chosen as
\cite{Ertl:2000si}
\begin{equation}
  \label{eq:dpabosc}
  \bar{C} = \bar{Y} - \inv{8} \bar{u}^2 + \inv{16}
  \bar{\chi}^2 \bar{u}'\ .
\end{equation}
Because the on-shell Casimir function is a constant it must be conformally
invariant. Thus a simple change of variables according to  \eqref{eq:ct1} and
\eqref{eq:ct2} leads to
\begin{align}
  \label{eq:genbosc}
  C & = e^{Q}\bigl(Y - \inv{8}
  u^2 + \inv{16} \chi^2 C_\chi \bigr)\ , \medsp
\label{eq:Cchi}
  C_\chi &= u' + \half{1} u Z\ .
\end{align}
Eliminating the auxiliary field $X^a$ by \eqref{eq:genxa}, the Casimir function
of the MFDS model of eq.\ \eqref{eq:mostgenaction2} becomes
(the special case of NDMFS \eqref{eq:simplifiedaction2} is found by setting $Q = Z = 0$)
\begin{equation}
  \label{eq:elimcas}
  \itindex{C}{MFDS} = e^{Q}\bigl( - \half{1} \partial^n \phi \partial_n \phi - \inv{8} u^2 -
  \half{1} \partial^n \phi (\chi \gthree \psi_n) + \inv{16} \chi^2 (u' +
  \half{1} u Z - \psi^n \psi_n) \bigr)\ .
\end{equation}

For explicit calculations the expressions are written more conveniently in terms of light-cone
coordinates (cf.\ Appendix \ref{sec:notation}). Assuming $X^{++} \neq 0$ one can introduce the Lorentz-scalars
\begin{align}
  \rho\bplus &= \frac{\chi^+}{\sqrt{|X^{++}|}}\ , & \rho\bminus &=
  \sqrt{|X^{++}|} \chi^-\ .
\end{align}
The solution of the MFS$_0$ model \eqref{eq:MFS0action} has been derived already in ref.\
\cite{Ertl:2000si} sect.\ 8,
the conformal transformation \eqref{eq:ct1}, \eqref{eq:ct2} of which yields
the general solution of MFS \eqref{eq:mostgenaction}. The strategy to
obtain in a straightforward way the general solution for a (g)PSM model 
consists in following the steps set out in ref.\ \cite{Kummer:1995qv} (cf.\
\cite{Grumiller:2002nm}). The final result is best parametrized in terms of
(almost-)Casimir-Darboux coordinates, which can be identified a posteriori. Indeed, introducing new gauge-potentials
$\mathcal{A}_I = (A_C,A_\phi,A_{++},A_{(+)},A_{(-)})$ that correspond to the target space
variables $\mathcal{X}^I = (C,\phi,X^{++},\rho\bplus,\rho\bminus)$ (cf.\ \eqref{eq:basicconf}
and \eqref{eq:gaugepotconf}), all $\mathcal{A}_I$ can be expressed in terms of the
$\mathcal{X}^I$ by the solution of their e.o.m.-s, except for
$A_C$. The e.o.m. of $A_C$ simply reads
\begin{equation}
  \extd C = 0
\end{equation}
and therefore we introduce a new integration function $F$:
\begin{align}
\label{eq:dFdefinition}
  \extd C &= 0 &&\Longrightarrow & \extd A_C &= 0 && \Longrightarrow & A_C &=
  - \extd F
\end{align}

Thus the solution is parametrized in terms of the target space variables
$\mathcal{X}^I$ and the free
function $F$.
Denoting by $\mathcal{V}$ the component of $P^{ab} = \mathcal{V} \epsilon^{ab}$
(cf.\ \eqref{eq:mostgensup})  
\begin{equation}
  \mathcal{V} = V + Y Z - \half{1} \chi^2 \Bigl( \frac{VZ + V'}{2u} +
  \frac{2 V^2}{u^3} \Bigr)\ ,
\end{equation}
the general analytic solution on a patch with $X^{++} \neq 0$ and $C = \mbox{const.} \neq 0$ can be written as
\begin{align}
\label{eq:sol11}
  \begin{split}
    \omega &= \frac{\extd X^{++}}{X^{++}} - e^Q \mathcal{V} A_\Upsilon -
   \frac{e^{Q}}{8 C}
   C_\chi \rho\bplus \extd \Upsilon \medsp
    &\quad - \frac{Z}{4} e^Q \bigl( \frac{u Z}{8C} \rho\bminus \rho\bplus
   \extd \phi  + \frac{1}{4} C_\chi \rho\bminus
    \rho\bplus \extd F - \frac{\sigma}{\sqrt{2} C} \rho\bminus \extd
   \Upsilon \bigr)\ ,
  \end{split}\medsp
\label{eq:sol12}
  \begin{split}
  X^{++} e_{++} &= - \extd \phi - e^Q X^{++} X^{--} A_\Upsilon - e^Q
   \frac{u}{16 C} \rho\bminus \rho\bplus \extd \phi\medsp
   &\quad + \frac{\sigma}{4 \sqrt{2}} \Bigl(\frac{e^Q}{C}
   (\rho\bminus + \frac{\sigma u}{2 \sqrt{2}}\rho\bplus) \extd
   \Upsilon - \rho\bplus \extd \rho\bplus \Bigr)\ ,
   \end{split}\medsp
\label{eq:sol13}
   \frac{e_{--}}{X^{++}} &=  - e^Q A_\Upsilon\ , \medsp
\label{eq:sol14}
     \sqrt{|X^{++}|} \psi_+ &= \frac{e^Q}{8} C_\chi \rho\bminus A_\Upsilon  + \frac{\sigma}{2 \sqrt{2}} \Bigl(  \extd
     \rho\bplus - e^Q \frac{u \sigma}{2 \sqrt{2} C} \bigl( \extd
     \Upsilon + \half{1} Z \Upsilon \extd \phi \bigr)
     \Bigr)\ , \medsp
\label{eq:sol1last}
    \frac{\psi_-}{\sqrt{|X^{++}|}} &= - \frac{e^Q}{8} C_\chi \rho\bplus
    A_\Upsilon  + e^Q \frac{\sigma}{2 \sqrt{2} C} \bigl( \extd
    \Upsilon  + \half{Z} \Upsilon \extd \phi \bigr)\ .
 \end{align}
Beside the abbreviation
$C_\chi$ in \eqref{eq:Cchi} a new variable and its gauge potential, namely ($\sigma = \mbox{sign} X^{++}$)
\begin{align}
  \label{eq:alambda}
   \Upsilon &= \rho\bminus - \frac{\sigma
  u}{2 \sqrt{2}} \rho\bplus\ , & A_\Upsilon &= \extd F + e^Q \frac {\sigma}{4 \sqrt{2} C^2} \Upsilon \extd
  \Upsilon\ ,
\end{align}
have been introduced. It should be noticed that in \eqref{eq:sol14} and
\eqref{eq:sol1last} half of the terms produced by $\Upsilon$
in $A_\Upsilon$ vanish due to the Grassmann property $(\rho\bplus)^2 =
(\rho\bminus)^2 = 0$.

In \eqref{eq:sol11}-\eqref{eq:sol1last} $X^{--}$ and $\Upsilon$ are dependent
variables according to eqs.\
\eqref{eq:genbosc} with $Y = X^{++} X^{--}$ at $C = \mbox{const.} \neq 0$ and
\eqref{eq:fermc}. Further arbitrary functions are $\phi$, $\extd F$ and the
fermionic $\rho\bplus$, $\rho\bminus$.

It is straightforward to check that the spinor $\tilde{c}$ defined
as ($\sigma = \mbox{sign} X^{++}$)
\begin{align}
\label{eq:fermc}
  \tilde{c} &= e^{\half{1} Q} \Upsilon
\end{align}
commutes\footnote{All commutators refer to its definition below \eqref{eq:gPSMaction}.} with everything but with itself. From the Schouten bracket
\begin{equation}
  \{ \tilde{c}, \tilde{c} \} = - 2 \sqrt{2} \sigma e^Q C
\end{equation}
it follows that for $C \equiv 0$ an additional fermionic Casimir function
$\tilde{c}$ arises. On a patch with $X^{++} = 0$, but $X^{--} \neq 0$ we can define an
analogous quantity $\hat{c}$ with $\rho\bminus$ and $\rho\bplus$ interchanged.
$\tilde{c} = \mbox{const.}$ relates $\rho\bplus$ and $\rho\bminus$ (cf.\
\eqref{eq:fermc}). Its associated gauge potential is (cf.\ \eqref{eq:dFdefinition}) $\tilde{A} = - \extd f$.
The general solution reads
\begin{align}
\label{eq:sol21}
  \begin{split}
    \omega &= \frac{\extd X^{++}}{X^{++}} - e^Q \mathcal{V} \extd F +
   e^{\half{1} Q} \frac{\sigma}{2 \sqrt{2}}
   C_\chi \rho\bplus \extd f \medsp
    &\quad - \frac{Z}{2} e^{\half{1}Q} \bigl( \rho\bminus \extd
   f  + e^{\half{1} Q} \frac{1}{8} C_\chi \rho\bminus
    \rho\bplus \extd F  \bigr)\ ,
  \end{split}\medsp
\label{eq:sol22}
  \begin{split}
  X^{++} e_{++} &= - \extd \phi- e^Q X^{++} X^{--} \extd F\medsp
  &\quad - \inv{2}
   \Bigl( \frac{\sigma}{2 \sqrt{2}} \rho\bplus \extd
   \rho\bplus + e^{\half{1}Q}
   (\rho\bminus + \frac{\sigma u}{2 \sqrt{2}}\rho\bplus) \extd
   f  \Bigr)\ ,
   \end{split}\medsp
\label{eq:sol23}
   \frac{e_{--}}{X^{++}} &=  - e^Q \extd F\ , \medsp
\label{eq:sol24}
     \sqrt{|X^{++}|} \psi_+ &= \frac{e^Q}{8} C_\chi \rho\bminus \extd F
     + \frac{\sigma}{2 \sqrt{2}} \Bigl(  \extd
     \rho\bplus + e^{\half{1}Q} u   \extd
     f \Bigr)\ ,
    \medsp
\label{eq:sol2last}
    \frac{\psi_-}{\sqrt{|X^{++}}|} &= - \frac{e^Q}{8} C_\chi \rho\bplus \extd F - e^{\half{1}Q}  \extd f\ .
\end{align}
Beside the anti-commuting constant $\tilde{c}$ the free functions of this solution are $\extd F$, $\extd
f$, $\phi$, $X^{++}$ and $\rho\bplus$. $X^{--}$ and $\rho\bminus$ are
dependent variables according to \eqref{eq:genbosc} with $C = 0$ and
\eqref{eq:fermc} with $\tilde{c} = \mbox{const.}$

For certain potentials \eqref{eq:bosonpot} also solutions with $X^{++} =
X^{--} = 0$ may appear, which can describe a ``supersymmetric ground-state''
\cite{Park:1993sd}. Then $\chi = 0$ and the discussion reduces to the pure
bosonic case (cf.\ e.g.\ ref.\ \cite{Grumiller:2003ad} for a situation where
such a solution appears).

\section{Coupling of Supersymmetric Test-Particle in\\ Minimal Field Supergravity}
\label{sec:sparticle}
To find the proper invariant coupling of a test-particle to supergravity seems
to be a hopeless task when one stays within the gPSM related MFS
model. On the other hand, in order
to study global properties for any of the solutions obtained above a ``super-geodesic'' is
needed. For a problem of this type, where a simple access to an invariant
expression is needed, the superfield approach is the method of choice. The
path of a super-particle is described by the map $\tau \rightarrow z^M(\tau)$ with coordinates\footnote{$x^m =
  x^m(\tau)$ and $\theta^\mu = \theta^\mu(\tau)$ are taken in this section
  without explicitly indicating the difference to the free variables $x$ and
  $\theta$ in the preceding sections.}
\begin{equation}
  \label{eq:wscoord}
  z^M = (x^m, \theta^\mu)\ .
\end{equation}
Holonomic indices are transformed into anholonomic ones according to
\begin{align}
\label{eq:indextransform}
  x^a &= e^a_m x^m\ , & \theta^\alpha &= \delta^\alpha_\mu \theta^\mu\ .
\end{align}
Due to the second equation \eqref{eq:indextransform} no separate notation
(cf.\ \eqref{eq:Achi})
for the components of $\theta^\mu$ is needed and
$\theta^\mu = (\theta^+, \theta^-)$.

The action of a super-particle with mass $m$ moving along the curve $z^M(\tau)$ may be written as
\cite{Casalbuoni:1976hx,Brink:1981nb,Siegel:1983hh,Galajinsky:1996rp,Knutt-Wehlau:1998gq}
\begin{equation}
  \label{eq:generalsp}
  \Stext{PP} = \intd{\diff{\tau}} \Bigl( g^{-1} \bigl(\dot{z}^M {E_M}^{++} \dot{z}^N
  {E_N}^{--} \bigr) + \half{m^2} g - m \dot{z}^M {E_M}^A \Gamma_A \Bigr)\ .
\end{equation}
It exhibits the well-known additional fermionic $\kappa$
symmetry \cite{Siegel:1983hh}:
\begin{equation}
\begin{split}
\label{eq:generalkappa}
 \delta_\kappa z^M {E_M}^a &= 0\medsp
 \delta_\kappa z^M {E_M}^+ &= - \bigl( m \kappa^+ + \sqrt{2}
  \frac{\kappa^-}{g} \dot{z}^M {E_M}^{++} \bigr) \medsp
 \delta_\kappa z^M {E_M}^- &= - \bigl( m \kappa^- + \sqrt{2}
  \frac{\kappa^+}{g} \dot{z}^M {E_M}^{--} \bigr) \medsp
 \delta_\kappa g &= -4 \dot{z}^M \bigl({E_M}^+ \kappa^- + {E_M}^- \kappa^+
 \bigr)
\end{split}
\end{equation}
The action \eqref{eq:generalsp} is a condensed expression which includes both, the
massless and the massive case. The limit $m
\rightarrow 0$ of \eqref{eq:generalsp} leads to the standard action of the
massless pointparticle, while the formula for the massive particle to be found in
most of the literature is
recovered by rescaling
$\tilde{g} = - m g$ and $\tilde{\kappa}^\pm = - m \kappa^\pm$.

In contrast to bosonic gravity the action \eqref{eq:generalsp} does not
contain the full super-line element
\begin{equation}
  \label{eq:susylineel}
  (ds)^2  = \extd z^M \otimes \extd x^N G_{NM} = 2 \extd z^M {E_M}^{++} \otimes
  \extd z^N {E_N}^{--} + 2 \extd z^M {E_M}^+ \otimes \extd z^N {E_N}^-\ .
\end{equation}
The standard super-particle \eqref{eq:generalsp} with $m = 0$ only considers the first
part of \eqref {eq:susylineel}, including bosonic anholonomic indices to be
summed over\cite{Casalbuoni:1976hx,Brink:1981nb,Siegel:1983hh}, which by itself is invariant under
supergravity transformations. We do not provide a detailed
comparison of the consequences of the two approaches \eqref{eq:generalsp} and \eqref{eq:susylineel} within this work. But it is important to notice
that even the case $m=0$ in \eqref{eq:generalsp} leads to different equations
of motion in the supersymmetry sector than the ones following from
\eqref{eq:susylineel}\footnote{What is meant by ``global'' properties of a
  solution is well-known to depend on the ``device'' by which
  (super-)geodesics are defined. Already in the purely bosonic case with
  non-vanishing torsion the use of ``geodesics'' (depending on Christoffel
  symbols only) or ``autoparallels'' (depending also on the contorsion) may lead
to different global properties.}.

In the standard gauge \eqref{eq:szb1}-\eqref{eq:szblast}
the connection $\Gamma^A$ reduces to the result in flat superspace
with the only non-vanishing components
\begin{align}
  \Gamma_+ &= \theta^-\ , & \Gamma_- &= \theta^+\ .
\end{align}
To explore the global structure of two-dimensional supergravity with this
super-particle the solutions \eqref{eq:sol11}-\eqref{eq:sol1last} and
\eqref{eq:sol21}-\eqref{eq:sol2last} must be inserted in
\eqref{eq:generalsp}. To this end the $\theta$-expansion of
\eqref{eq:generalsp} must be calculated explicitly. After some super-algebra the first term of
\eqref{eq:generalsp} (relevant for the massless super-particle) takes the form
\begin{equation}
  \label{eq:mlessexp}
  \begin{split}
    g^{-1} \bigl(\dot{z}^M {E_M}^{++} \dot{z}^N
  {E_N}^{--} \bigr) &= g^{-1} \Bigl(\dot{x}^m e_m^{++} + \sqrt{2}
  \dot{\theta}^+ \theta^+ + 2 \sqrt{2}  \dot{x}^m \underline{\psi}_m^+ \theta^+ +
  \theta^{-} \theta^{+} \underline{A} \dot{x}^m e_m^{++} \Bigr) \medsp
  &\quad \Bigl(\dot{x}^n e_n^{--} +  \sqrt{2}
  \dot{\theta}^- \theta^- +  2 \sqrt{2} \dot{x}^n \underline{\psi}_n^- \theta^- +
  \theta^{-} \theta^{+} \underline{A} \dot{x}^n e_n^{--} \Bigr)\ ,
  \end{split}
\end{equation}
while the Wess-Zumino contribution becomes
\begin{equation}
  \label{eq:GammaAexp}
  \dot{z}^M {E_M}^A \Gamma_A = \dot{\theta}^+ \theta^- + \dot{\theta}^-
  \theta^+ + \dot{x}^m \underline{\psi}_m^+ \theta^- + \dot{x}^m \underline{\psi}_m^- \theta^+ +
  \dot{x}^m \tilde{\underline{\omega}}_m \theta^- \theta^+\ .
\end{equation}
When inserting the classical solution for $\underline{A}$ (eq.\ \eqref{eq:psauxfield2})
and the explicit expression for the dependent spin-connection $\tilde{\underline{\omega}}$
(cf.\ eq.\ \eqref{eq:tildeomega}) in \eqref{eq:mlessexp} and \eqref{eq:GammaAexp},
the action of the supersymmetric test-particle
\eqref{eq:generalsp} is parametrized in terms of the zweibein $e_m$, the
gravitino $\underline{\psi}_m$, the dilaton field $\underline{\phi}$ and the dilatino
$\underline{\chi}$. By means of the identification \eqref{eq:genequiv} and \eqref{eq:psi1}
the action \eqref{eq:generalsp} turns into a function of $e_m$, $\psi_m$, $\phi$
and $\chi$:
\begin{align}
  \label{eq:PPwithGPSM}
    \Stext{PP} &= \intd{\diff{\tau}} \Bigl\{ g^{-1} A^{++} A^{--} +
    \half{m^2} g - m (B^{+-} + B^{-+})  \Bigr\} \medsp
  \begin{split}
  \label{eq:A++}
    A^{++} &= \dot{x}^m e_m^{++}
    \Bigl( 1 + \half{1} Z \chi^- \theta^+ - \half{1}\theta^2 \bigl(\half{1} u
    Z + \half{1} u' - \inv{16} Z' \chi^2 \bigr) \Bigr) \medsp
    &\quad + \sqrt{2} \dot{\theta}^+ \theta^+ + 2 \sqrt{2}  \dot{x}^m
    \psi_m^+ \theta^+
  \end{split} \medsp
  \label{eq:B+-}
  B^{+-} &= \dot{\theta}^+ \theta^- + \inv{4} \theta^2 \dot{x}^m
    \tilde{\omega}_m + \dot{x}^m \psi_m^+ \theta^- + \frac{Z}{4 \sqrt{2}} \dot{x}^m
    e_m^{++} \chi^- \theta^-
\end{align}
Here $A^{--}$ and $B^{-+}$ are defined through \eqref{eq:A++} and
\eqref{eq:B+-} by the interchange of all \emph{explicit} anholonomic indices
$+ \rightarrow -$, $- \rightarrow +$. 
\subsection{Gauge Choice}
When the supersymmetric test-particle moves on the
 supergravity background, the zweibein and the gravitino in
 \eqref{eq:PPwithGPSM} are replaced by their classical solutions
\eqref{eq:sol12}-\eqref{eq:sol1last} or
\eqref{eq:sol22}-\eqref{eq:sol2last} resp. 
In principle, ``super-geodesics'' could then be obtained from variation of
\eqref{eq:generalsp} or \eqref{eq:PPwithGPSM}, respectively. This task simplifies
considerably when an appropriate gauge-fixing is used:
\begin{enumerate}
\item \label{enum:gf1} The solutions from MFS depend, among others, on
the variable $X^{++}$, which is not present in superspace. The supersymmetric
test-particle being manifestly invariant under
local Lorentz transformations, we can eliminate
this dependence by a (finite) local Lorentz transformation . Thus on any patch with $X^{++} \neq 0$ we can fix
its value to $X^{++} = 1$ or $X^{++} = -1$, depending on the sign of the
original configuration. For $X^{++} = 0$ we have to parametrize the solution
analogously in terms of $X^{--}$ (cf.\ section \ref{sec:solution}).
\item \label{enum:gf2}$\kappa$-symmetry can be used to gauge one of the fermionic variables to
  a constant \cite{Lindstrom:1989qa}. It turns out that $\dot{\theta}^- \equiv
  0$ is the preferable choice for $X^{++} \neq 0$.
\item \label{enum:gf3}It has been argued in ref.\ \cite{Strobl:1999zz} that the classical
  solution \eqref{eq:sol11}-\eqref{eq:sol1last} is equivalent to the
  corresponding solution of the purely bosonic model up to local supersymmetry
  transformations. Thus locally all fermionic target-space degrees
  of freedom could be gauged away: $\rho\bplus \equiv 0$, $\rho\bminus \equiv 0$ and consequently
  $\psi_\pm \equiv 0$. One might ask whether this zero fermion (ZF) gauge is accessible and
  allowed.

  Concerning the question whether the ZF gauge is allowed, one should consult
  the situation in the purely bosonic case. There the line element, after elimination of $Y = X^{++}
  X^{--}$ by means of the Casimir constant, is determined by
  two arbitrary functions $F$ and $\phi$. The ``gauge'' $\extd
  F = 0$ is forbidden by the requirement of non-singular gravity, namely that the
  determinant of the metric should be different from zero. It would be
  natural, although not strictly necessary, to transfer these arguments directly
  to supergravity, i.e.\ demanding a non-singular super-determinant. However,
  as can be seen from \eqref{eq:superdet}, the vanishing of that determinant
  is controlled by its bosonic contributions. Thus within this line of arguments the ZF
  gauge is allowed.

  Typically the accessibility of a gauge is more difficult to
  answer than the question whether it is allowed. Beside the mere mathematical
  challenge to describe finite gauge transformations, accessibility involves
  subtle physical questions as well. For MFS the mathematical aspect found a
  definite answer \cite{Strobl:1999zz}:  By means of a finite gPSM
  gauge transformation any solution can be brought to ZF gauge\footnote{The explicit proof
  had been performed in ref.\ \cite{Strobl:1999zz} for MFS$_0$ only, but it
  generalizes straightforwardly to MFS by the use of the conformal
  transformations \eqref{eq:ct1} and \eqref{eq:ct2}.}.

  The
  physical aspect is more involved. Indeed, under a finite transformation not only the
  the specific solution of the field equations itself, but also the ``device''
  defining the (super-)geodesics (e.g.\
  the
  super-pointparticle action) must be transformed. This last
  step can be omitted if and only if the
  \emph{new} solution together with the \emph{old} device turns out to have the
  same physics as the \emph{new} solution with some ``invariant''
  device\footnote{\label{foot:ts1}A simple example is a wave packet solution
  in classical field
  theory. Clearly the solution breaks (global) rotation symmetry as the wave
  packet moves in a certain direction. In an ``invariant'' system of
  detectors, the latter are (or can be) arranged in a rotationally symmetric way. Then
  physics (measurement of the wave packet) remains the same.}.
 This does not necessarily imply that the solution
  does not transform at all, but solely that the two systems are physically
  (albeit not mathematically) equivalent. This is often the case for broken
  \emph{bosonic} symmetries, where states (field configurations) exhibit such
  a degeneracy. On the other hand, some well-known symmetries do not
  allow the simplification of using the old device: The conformal transformation
  discussed in section \ref{sec:four} is a bosonic example for that.
  In the present context it is important that \emph{broken} supersymmetry does
  not allow the above shortcut as well. Indeed, breaking of supersymmetry
  never leads to an equivalent class of states with the same physical
  properties\footnote{In contrast to the example in footnote \ref{foot:ts1} broken
  supersymmetry acting on a bosonic wave packet produces fermions. Then
  it is obviously relevant whether the detector has been transformed too. If
  this is the case it would still register ``bosons'' although it
  would receive fermionic contributions as well.}. But, as
  may be checked easily by inserting any of the solutions of Section
  \ref{sec:solution}, including the one considered in ref.\
  \cite{Strobl:1999zz}, into the supersymmetry transformations
  \eqref{eq:gPSMtransf3} and \eqref{eq:gPSMtransflast}, many of them
  break at least half of the supersymmetries (cf.\ ref.\ \cite{Park:1993sd}
  and the systematic study of supersymmetric solutions in ref.\ \cite{Bergamin:2003mh}). Therefore, with respect to the transformation proposed in
  ref.\ \cite{Strobl:1999zz} the device must be transformed as well.  Thus we
  expect that in the generic case the global properties of the solutions of
  Section \ref{sec:solution} do depend on fermionic background fields, if
  these fields cannot be transformed away by means of \emph{unbroken} supersymmetries.
  
  Despite the problems of physical accessibility of the ZF gauge we will, for simplicity, restrict our
  analysis below to this specific class of solutions. This choice also correlates with the observation
  that classical field equations usually possess solutions with vanishing fermion
  fields\footnote{There are, of course, counter-examples, e.g.\ the solution
  \eqref{eq:sol21}-\eqref{eq:sol2last} for $\tilde{c} \neq 0$ in \eqref{eq:fermc}.}.
  Inserting it into the super-determinant
  \eqref{eq:superdet} yields a non-trivial result, namely
  \begin{equation}
    \label{eq:superdetspecial}
    E = e \bigl(1 - \half{1} \theta^- \theta^+ ( u Z + u') \bigr)\ .
  \end{equation}
  For non-singular gauges --in the usual sense-- the superdeterminant is
  non-vanishing and
  does \emph{not} reduce to the purely bosonic result, although the
  dilatino and the gravitino of the background have been gauged away, because the fermionic partner
  $\theta^+(\tau)$ of the bosonic geodesic $x^m(\tau)$ survives.

  Besides the drastic simplification of the pointparticle action, this gauge
  is very convenient also from the technical point of view,  as it permits an
  easy application for both solutions
  \eqref{eq:sol11}-\eqref{eq:sol1last} and
  \eqref{eq:sol21}-\eqref{eq:sol2last} derived in the previous section.
  It should be kept in mind, though, that the ZF gauge is
  problematic. Nevertheless, for a first cursory exploration of
  super-geodesics derived from \eqref{eq:PPwithGPSM}-\eqref{eq:B+-} it
  certainly is a convenient starting point.
\item \label{enum:gf4}In the bosonic sector an
  Eddington-Finkelstein like gauge is the most convenient one \cite{Grumiller:2002nm}. To this end the
  world-sheet coordinates $x^m$ are chosen such that the remaining target space
  variables $F(x)$ and $\phi(x)$ describe the trivial embedding
\begin{align}
\label{eq:EFgauge}
 F(x) &\equiv
  x^0(\tau) = F(\tau)\ , & \phi(x) &\equiv x^1(\tau) = \phi(\tau)\ .
\end{align}
  These will be our
  bosonic coordinates in the following.
\end{enumerate}
The pointparticle action \eqref{eq:PPwithGPSM} with the solution
\eqref{eq:sol11}-\eqref{eq:sol1last} in the gauge choice as proposed in paragraphs
\ref{enum:gf1}-\ref{enum:gf4} above, together with \eqref{eq:EFgauge} simplifies to
\begin{gather}
  \label{eq:PPgf}
  \Stext{PP} = \intd{\diff{\tau}} \bigl( \Stext{PP}^{\mbox{\tiny{bosonic}}} +
  \Stext{PP}^{\mbox{\tiny{SUSY}}} + \frac{m^2}{2} g \bigr)\ , \medsp
\label{eq:PPgfbosonic}
  \Stext{PP}^{\mbox{\tiny{bosonic}}} = g^{-1} e^Q \dot{F} \bigl( \dot{\phi} +
  \xi(\phi) \dot{F} \bigr)\ , \medsp
\label{eq:PPgfSUSY}
\begin{split}
  \Stext{PP}^{\mbox{\tiny{SUSY}}} &= g^{-1} \Bigl[ \theta^- \theta^+ \mu(\phi)
  \cdot e^Q \dot{F} \bigl( \dot{\phi} + \xi(\phi) \dot{F} \bigr) - \sqrt{2}
  \dot{\theta}^+ \theta^+ (\dot{\phi} + \xi \dot{F}) 
  \Bigr] \medsp
  &\quad - m \Bigr[\theta^- \dot{\theta}^+ + \theta^- \theta^+ \bigl( \dot{F}
  \bigl( Z \cdot \xi(\phi) + \xi'(\phi) \bigr) + Z \cdot \dot{\phi} \bigr) \Bigr]\ .
\end{split}
\end{gather}
The dependence on the dilaton $\phi$ occurs in the coefficient
of bosonic torsion $Z$ and in two functions $\xi(\phi)$ and $\mu(\phi)$
defined as
\begin{align}
\label{eq:ximudef}
  \xi(\phi) &=  C + \inv{8} e^Q u^2\ , & \mu(\phi) &= - u \cdot Z - u'\ .
\end{align}
\subsection{Orbits of the Massless Pointparticle}
To explore the global structure of a certain gravitational background the
equations of motion from
\eqref{eq:PPgf}-\eqref{eq:PPgfSUSY} have to be derived. To this end the
variation with respect to the super-coordinates $\{z^M\} =
\{F,\phi,\theta^+ \}$ has to be calculated. It is convenient to re-parametrize the curve
$z^M(\tau)$ in such a way that $\dot{g} \equiv 0$. Then the variations can be
written as:
\begin{align}
\label{eq:geodv}
  \begin{split}
    g \cdot \delta_F \Stext{PP} &= \derfrac{\tau} \biggl\{ - e^Q (\dot{\phi} +
  2 \xi \dot{F})(1 + \theta^- \theta^+ \mu) + \sqrt{2} 
  \dot{\theta}^+ \theta^+ \xi \medsp
  &\quad \phantom{\derfrac{\tau} \biggl\{} + m g \theta^- \theta^+ (Z \xi + \xi') \biggl\}
  \end{split}\medsp
\label{eq:geodphi}
  \begin{split}
    g \cdot \delta_\phi \Stext{PP} &= e^Q (1 + \theta^- \theta^+ \mu)
    \bigl((\xi' + Z \xi)
    \dot{F}^2 - \ddot{F}\bigr) + e^Q \theta^- \theta^+ \mu' \xi \dot{F}^2 -
    e^Q \theta^-
    \dot{\theta}^+ \mu \dot{F} \medsp
    &\quad - \sqrt{2} \dot{\theta}^+ \theta^+ \xi' \dot{F} + \sqrt{2}
    \ddot{\theta}^+ \theta^+ - m g \bigl(\theta^- \theta^+ \dot{F} (Z' \xi
    + Z \xi' + \xi'') - \theta^- \dot{\theta}^+ Z \bigr)
  \end{split} \medsp
\label{eq:geodthetaplus}
  \begin{split}
    g \cdot \delta_{\theta^+} \Stext{PP} &= - e^Q \theta^- \mu \dot{F}(\dot{\phi}
    + \xi \dot{F}) + \sqrt{2} \theta^+ (\ddot{\phi} + \xi' \dot{\phi}
    \dot{F} + \xi \ddot{F}) + 2 \sqrt{2} \dot{\theta}^+ (\dot{\phi} + \xi
    \dot{F}) \medsp
    &\quad + mg \theta^- \bigl(\dot{F}(Z \xi + \xi') + \dot{\phi} Z \bigr)  
  \end{split} 
\end{align}
Equation \eqref{eq:geodv} is a total derivative as $\derfrac{F}$ is a Killing
field. The expression in the curly brackets corresponds to the related
constant of motion.
To solve \eqref{eq:geodphi} and \eqref{eq:geodthetaplus} with the constant of
motion \eqref{eq:geodv} in full generality is a daunting task which we do not attempt in the
present work. For illustrative purposes we find it sufficient to consider
special cases with some physical relevance.
The massless particle ($m = 0$) together with its supersymmetric orbit
$\theta^+(\tau)$ already allows the discussion of different physically
interesting analytic solutions. In the cases to be treated below further
restrictions will be made:
\renewcommand{\labelenumii}{(\arabic{enumi}\alph{enumii})}
\begin{enumerate}
\item\label{item:MKground} ``Minkowski ground state models'': Within
  bosonic theories of gravity a special subset is determined by the condition
  that at $C = 0$ in \eqref{eq:bosonicC} the metric of these theories
  reduces to the one of Minkowski space
  \cite{Katanaev:1996bh,Grumiller:2002nm}. This implies a relation between the
  functions $Z(\phi)$ and $V(\phi)$ in \eqref{eq:bosonpot}, which for
  supergravity becomes just the condition $\mu = 0$, i.e.\ the relation
  \begin{equation}
    \label{eq:muzero}
    \frac{u'}{u} = - Z
  \end{equation}
  between the prepotential and the function which determines non-vanishing
  bosonic torsion. Spherically reduced gravity from $d = 4$ belongs to this
  class with ($l$ is an arbitrary constant)
  \begin{align}
    \label{eq:SRG}
    \itindex{Z}{SRG} &= - (2 \phi)^{-1}\ , & \itindex{u}{SRG} = -l
    \sqrt{\phi}\ ,
  \end{align}
  but also more general models which are asymptotically flat, if $u(\infty)
  \rightarrow \infty$.
\item The (bosonic) light-like directions
  $\Stext{PP}^{\mbox{\tiny{bosonic}}} = 0$ correspond 
  to especially simple solutions. They are characterized by
  \begin{enumerate}
  \item\label{item:Fdot} $\dot{F} = 0$,
  \item\label{item:Fphi} $\dot{\phi} + \xi \dot{F} = 0$.
  \end{enumerate}
\end{enumerate}
\subsubsection{Minkowski Ground-State Models}
The solution
of the purely bosonic model with $\mu = 0$ is regarded as a given input in
this subsection. Actually, the interesting
cases will be covered by solutions of the type (\ref{item:Fdot}) and
(\ref{item:Fphi}) above.

For $\mu = 0$ and $m = 0$ the variation \eqref{eq:geodthetaplus} vanishes if
\begin{equation}
\label{eq:mgrone}
   \partial_\tau(\dot{\phi} + \xi \dot{F}) \theta^+ + 2 (\dot{\phi} + \xi
   \dot{F}) \dot{\theta}^+ = 0
\end{equation}
holds. The solutions can be classified as follows:
\renewcommand{\labelenumi}{(\Alph{enumi})}
\renewcommand{\labelenumii}{(\Roman{enumii})}
\renewcommand{\theenumi}{\Alph{enumi}}
\renewcommand{\theenumii}{.\Roman{enumii}}
\begin{enumerate}
\item\label{item:A} $\dot{\phi} + \xi \dot{F} = 0$. This is the special case
  $\mu = 0$ of
  (\ref{item:Fphi}) above and will be discussed together with the generic
  solutions of this type below.
\item\label{item:B} $\dot{\theta}^+ = \half{1} \partial_\tau \log (\dot{\phi} + \xi
  \dot{F}) \theta^+$\ .
\end{enumerate}
The general solution of \eqref{item:B} is
\begin{equation}
  \label{eq:MGSMsolution}
  \theta^+ = \inv{\sqrt{|\dot{\phi} + \xi \dot{F}|}} \lambda\ ,
\end{equation}
where $\lambda$ is an arbitrary constant spinor. The space of anti-commuting variables of this
class of solutions can be parametrized by the two constant spinors $\lambda$
and $\theta^-$. In a bosonic superfunction $A$ with body $A_B$ and soul $A_S$
therefore the decomposition
\begin{equation}
  \label{eq:thetalambdadecomp}
  A(\tau) = A_B(\tau) + A_S(\tau) = A_B(\tau) + \theta^- \lambda a(\tau)
\end{equation}
can be introduced. Here $A_B(\tau)$ and $a(\tau)$ are ordinary bosonic functions.

As \eqref{eq:MGSMsolution} does not depend on $\theta^-$ any
$\tau$-derivative of $\theta^+$ is again proportional to $\theta^+$ (or zero), which
especially means that $\theta^+(\tau)$ has no simple zeros. The singularity in
\eqref{eq:MGSMsolution} corresponds to a light-like direction \eqref{item:A}
of the bosonic line element.

Evaluating the variations \eqref{eq:geodv} and \eqref{eq:geodphi} with the
solution \eqref{eq:MGSMsolution} they simplify to
\begin{align}
\label{eq:mgrthree}
  \partial_\tau\bigl(e^Q(\dot{\phi} + 2 \xi \dot{F})\bigr) &= 0\ , \medsp
\label{eq:mgrfour}
  (\xi' + Z \xi) \dot{F}^2 - \ddot{F} &= 0\ ,
\end{align}
which are the relations from the purely bosonic model. The motion of $\theta^+$ is determined up to
the initial value according to \eqref{eq:mgrone}, i.e.\ up to the numerical value of
$\lambda$ in \eqref{eq:MGSMsolution}. Due to absence of the
$\theta$'s in \eqref{eq:mgrthree} and \eqref{eq:mgrfour}, the evolution of
$\phi$ and $F$ will not depend on
the fermionic variables. But as all bosonic quantities
must be regarded as (commuting) superfunctions resp.\ supernumbers, a soul can still be introduced
in them by an appropriate choice of the initial values for these fields. An example of this type is evaluated below.
\subsubsection{Light-like Solution (2a) with $\mathbf{\mu \neq 0}$}
The vanishing of  $\eqref{eq:geodthetaplus}$ relates the motion in the direction
$\theta^+$ to $\dot{\phi}$ and $\ddot{\phi}$:
\begin{equation}
  \label{eq:dthetaplus}
  \dot{\theta}^+ = - \half{1} \frac{\ddot{\phi}}{\dot{\phi}} \theta^+\ .
\end{equation}
Again the evolution of $\theta^+(\tau)$ does not depend on $\theta^-$ and the
general solution
\begin{equation}
  \label{eq:thetaplusoneone}
  \theta^+(\tau) = (\dot{\phi})^{- \inv{2}} \lambda
\end{equation}
with an arbitrary constant spinor $\lambda$ is an immediate consequence. The
space of anti-commuting coordinates
is again two-dimensional and may be parametrized by $\left(\theta^-, \lambda \right)$.

As a consequence of
\eqref{eq:thetaplusoneone} the term $\propto \dot{\theta}^+ \theta^+$ in the action
\eqref{eq:PPgfSUSY} vanishes. Thus the complete action \eqref{eq:PPgf} is
identically zero for $\dot{F} = 0$ and these orbits are not only null
directions of the bosonic part of the action, but of the whole super-particle
action \eqref{eq:PPgf}.

In terms of the solution \eqref{eq:thetaplusoneone} eq.\ $\eqref{eq:geodphi}$
vanishes identically, whereas the constant of motion \eqref{eq:geodv} can be brought
into the form
\begin{equation}
  \label{eq:dphi}
  e^Q (1 + \theta^- \theta^+ \mu) \dot{\phi} = k\ .
\end{equation}
Here $k$ is some constant super-number. After some straightforward
super-algebra the $\tau$ derivative of \eqref{eq:dphi} is found as
\begin{equation}
  \label{eq:ddphi}
  \ddot{\phi} = - \dot{\phi}^2 (Z + \half{1} \theta^- \theta^+ \mu Z +
  \theta^- \theta^+ \mu')\ .
\end{equation}
The body of this equation is seen to yield
the correct light-like geodesic $\ddot{\phi} = - Z \dot{\phi}^2$. Inserting
this relation into \eqref{eq:dthetaplus} (the terms $\propto \mu$ in
\eqref{eq:ddphi} vanish due to $(\theta^+)^2 = 0$) leads to 
\begin{align}
\label{eq:dthetaplus2}
  \dot{\theta}^+ &= \half{1} Z \theta^+ \dot{\phi}\ .
\end{align}
As $Z = d Q /d \phi$ (cf.\ \eqref{eq:bosonicC}) eq.\ \eqref{eq:dthetaplus2} can be transformed into a
total derivative and the general solution for $\theta^+$
\eqref{eq:thetaplusoneone} can be expressed alternatively as
\begin{equation}
\label{eq:thetaplusone}
  \theta^+ (\tau) = e^{\half{Q}} \lambda\ .
\end{equation}
The constant spinor $\lambda$ appearing in this solution is the same as the one in \eqref{eq:thetaplusoneone}.

To solve equation
\eqref{eq:dphi} a decomposition  according to \eqref{eq:thetalambdadecomp} is necessary. Using the relations
\begin{gather}
  \begin{align}
    k &= k_B + k_S\ , & \partial_\tau k_B &= \partial_\tau k_S = 0\ ,
  \end{align}\medsp
  e^{Q(\phi)} = e^{Q(\phi_B)}(1 + Z(\phi_B) \phi_S)\ ,
\end{gather}
body and soul of equation \eqref{eq:dphi} become
\begin{gather}
\label{eq:dphibody}
  e^{Q(\phi_B)} \dot{\phi}_B = k_B\ , \medsp
\label{eq:dphisoul}
  e^{Q(\phi_B)} \dot{\phi}_S + k_B Z(\phi_B) \phi_S + k_B \theta^- \theta^+
  \mu (\phi_B) = k_S\ ,
\end{gather}
where in eq.\ \eqref{eq:dphisoul} $\dot{\phi}_B$ has been eliminated by means of
\eqref{eq:dphibody}. The latter equation is equivalent to
the one of the bosonic model, but the value of the complete
$\phi(\tau)$ receives contributions from $\phi_S$ as well.
With the souls 
$\phi_S = \theta^- \lambda \varphi(\tau)$ and $k_S = \theta^- \lambda
\tilde{k}$  for $\phi$ and $k$ eq.\
\eqref{eq:dphisoul} becomes
\begin{equation}
  \label{eq:dvarphi}
  e^{Q(\phi_B)} \dot{\varphi} + k_B Z(\phi_B) \varphi + k_B e^{\frac{Q(\phi_B)}{2}} 
  \mu (\phi_B) = \tilde{k}\ .
\end{equation}
For a complete set of initial values
for super-coordinates $z^M (\tau = 0)$ and for the constant of motion $k$, eqs.\ \eqref{eq:thetaplusone},
\eqref{eq:dphibody} and \eqref{eq:dvarphi} uniquely determine the evolution of
the two dynamical variables $\phi$ and $\theta^+$. Certainly, eqs.\
\eqref{eq:dphibody} and \eqref{eq:dvarphi} cannot be solved in
general. In certain cases, however, a simple solution can be obtained:
\begin{itemize}
\item Due to \eqref{eq:dthetaplus2} the purely bosonic solution corresponds to
  a special choice of initial values, namely $\theta^+ = 0$. Indeed, if
  $\theta^+ = 0$ for any $\tau = \tau_0$, all $\tau$-derivatives on $\theta^+(\tau_0)$ vanish
  as well and consequently $\theta^+$ is zero everywhere. In this case the
  constant of motion has vanishing soul as well.
\item Models with vanishing bosonic torsion ($Z = 0$) possess a simple solution with non-trivial fermionic sector. As both $\theta^+$ and
  $\theta^-$ are constant in this case eq.\ \eqref{eq:dphi} becomes a total
  derivative (cf.\ eq.\ \eqref{eq:ximudef}):
  \begin{align}
    \label{eq:zzerosol1}
    \derfrac{\tau}(\phi - \theta^- \theta^+ u) &= k &&\Rightarrow& \phi -
    \theta^- \theta^+ u - c &= k \tau
  \end{align}
  Both $k$ and $c$ are supernumbers, where $c$ is determined by the initial
  value of $\phi$ as
  \begin{equation}
    c = \phi_0 - \theta^- \theta^+ u(\phi_0)\ .
  \end{equation}
  The explicit solution is
  \begin{align}
    \phi_B &= k_B \tau + \phi_0\ ,\medsp
    \phi_S &= \theta^- \theta^+ \bigl( u(\phi_B) -
    u(\phi_0) \bigr) + k_S \tau\ ,
  \end{align}
  where it has been assumed that the initial value of $\phi$ is bosonic:
  $\phi_0 = \phi_{B0}$. This class of solutions is especially interesting as
  it determines the null-directions of a massless super pointparticle on a
  background described by the model of Howe \cite{Howe:1979ia}.
\item As an example with non-vanishing bosonic torsion we consider spherically reduced gravity from four
  dimensions, i.e.\ a combination of \eqref{item:Fdot} with $\mu = 0$ and \eqref{eq:SRG}. As $\dot{\phi} + \xi
  \dot{F}\neq 0$ here, the situation is given by \eqref{item:B} above.
  Inserting \eqref{eq:SRG} into \eqref{eq:dphibody} and
  \eqref{eq:thetaplusone} and after integration of the former relation, one
  finds after a simple redefinition of $(1/2) (k_B \tau + c_B) \rightarrow \tau$
  \begin{align}
\label{eq:srgsol1}
    \phi_B &= \tau^2 \ , & \theta^+(\tau) &=
    \frac{\lambda}{\tau}\ .
  \end{align}
  The determination of the soul of $\phi$ simplifies in this special case, as
  $\mu(\phi) \equiv 0$. The integrating factor of the differential equation
  \eqref{eq:dvarphi} is given by
  \begin{equation}
    \rho(\tau) = \inv{2 \tau}\ , 
  \end{equation}
  which yields the last dynamical quantity as
  \begin{equation}
    \phi_S = \theta^- \lambda \tau (c_S + \tilde{k} \frac{2 \tau - c_B}{k_B})\ .
  \end{equation}
  Obviously the value of $\phi_S(\tau)$ does not depend on the evolution of
  $\theta^+(\tau)$ in \eqref{eq:srgsol1}. This is a consequence of $\mu = 0$ and, as already
  discussed above, holds in any
  Minkowski ground-state theory. Therefore, with the special choice $c_S =
  \tilde{k} = 0$ the soul of $\phi$ vanishes for all $\tau$, but of course, one is not
  forced to choose the integration constants like that. Nevertheless, a non-trivial coupling of
  the bosonic variable $\phi$ to the $\theta$ variables is somehow
  artificial. On the other hand, $\theta^+(\tau)$ does depend on the behavior
  of $\phi_B$ and thus even in this simple example the bosonic and the fermionic part of the pointparticle do
  not decouple from each other. The
  evolution of $\theta^+(\tau)$ shows a singularity at $\tau = 0$, a
  point where with our choice of $\tau$ the singularity of the Schwarzschild
  black hole at $\phi = 0$ is encountered.
\end{itemize}
\subsubsection{Light-like Solution (2b) with $\mathbf{\mu \neq 0}$}
Eq.\ \eqref{eq:geodthetaplus} vanishes trivially in this case, while eqs.\
\eqref{eq:geodv} and \eqref{eq:geodphi} (except for $\xi = 0$) are related by
\begin{equation}
  \mbox{\eqref{eq:geodv}} = \xi \cdot \mbox{\eqref{eq:geodphi}}\ .
\end{equation}
The remaining only differential equation reads
\begin{equation}
\label{eq:dphi2}
  e^Q (1 + \theta^- \theta^+ \mu ) \dot{\phi} + \sqrt{2} \xi
  \dot{\theta}^+ \theta^+ = k\ .
\end{equation}
If we had still $\dot{\theta}^+ \propto \theta^+$, eq.\ \eqref{eq:dphi2} does reduce to
\eqref{eq:dphi}, which, of course, is expected to hold for the body of the two
equations. Thus the solution
\eqref{eq:dthetaplus} is one of the solutions for \eqref{eq:dphi2}, with the
same $\phi(\tau)$ as derived in the previous section (notice however the
different behavior of $F$).

Nevertheless, \eqref{eq:dphi2} allows solutions
with $\dot{\theta}^+$ not proportional to $\theta^+$. This does not lead to a
non-vanishing action \eqref{eq:PPgf}, as the latter is proportional to $\dot{\phi} +
\xi \dot{F}$. Again one observes that any bosonic null-direction is a
null-direction of the whole super-particle.

To discuss the general solution the split into soul and body is made again:
\begin{align}
  e^{Q(\phi_B)} \dot{\phi}_B &= k_B \medsp
  e^{Q(\phi_B)} \dot{\phi}_S + k_B Z (\phi_B) \phi_S + k_B \theta^- \theta^+
  \mu(\phi_B) + \sqrt{2} \xi(\phi_B) \dot{\theta}^+ \theta^+ &= k_S
\label{eq:notrivialsoul}
\end{align}
By introducing the general decomposition for $\theta^+$
\begin{align}
  \theta^+(\tau) &= f(\tau) e^{\half{Q}} \lambda + g(\tau) \theta^-\ ,
  & f(0) &= 1\ , &
  g(0) &= 0\ ,
\end{align}
and using again $\phi_S(\tau) = \theta^- \lambda \varphi(\tau)$, \eqref{eq:notrivialsoul} becomes
\begin{multline}
  e^{Q(\phi_B)} \dot{\varphi} + k_B Z (\phi_B) \varphi + k_B e^{\half{1}
  Q(\phi_B)} \mu (\phi_B) f(\tau) \medsp
+ \sqrt{2} \xi(\phi_B) ( e^{\half{1}
  Q(\phi_B)} (\dot{g}(\tau) f(\tau) - \dot{f}(\tau) g(\tau)) -    e^{- \half{1}
  Q(\phi_B)} \half{1}k_B Z f(\tau)g(\tau) ) = \tilde{k}\ .
\end{multline}
Obviously the system is underdetermined, because three free functions $\varphi$,
$f$ and $g$  obey one single first order differential equation. The special
situation of $\mu = 0$ follows straightforwardly by setting this function to zero in
all equations of this subsection. 

\section{Conclusion and Outlook}

The formulation of supergravity as a superfield (dilaton) theory is
well-known since the seminal works of Howe \cite{Howe:1979ia} and Park, Strominger
\cite{Park:1993sd}. This approach uses the second order formalism with
vanishing bosonic torsion. Unless the dilaton superfield is non-dynamical, this
field appears in a second order action, which applies to most models
with a bosonic potential of direct physical
relevance like Einstein gravity, spherically reduced from D dimensions.
Also the string inspired CGHS model \cite{Callan:1992rs} (the formal limit D$\rightarrow\infty$) belongs to this class.

The complicated structure of the equations of motion in a second order
formalism, together with the large number of (auxiliary) field variables
in superfield supergravity, probably has been the main reason, why
to this day no full solution of generic 2d supergravity with dynamical
dilaton \cite{Park:1993sd} has been published\footnote{Cf.\ the comment below
  eq.\ (50) in \cite{Park:1993sd}.}.

Our present work closes this lacuna. Also for the first time -- we
believe -- a manageable treatment is provided for {}``supergeodesics'',
the motion of a super-point particle in a very general supergravity
background solution which, nonetheless, is described by a minimal set
of fields.

In order to achieve these results -- which by far do not exhaust the
list of further applications -- one relies heavily upon the knowledge
gained in first order 2d bosonic gravity \cite{Kummer:1992rt,Haider:1994cw} and
the more general concept of Poisson-sigma models \cite{Schaller:1994uj,Schaller:1994es}. 

Within this approach not only in bosonic gravity many (without interaction
with matter: essentially all) classical and quantum problems have
found a complete solution \cite{Grumiller:2002nm}, but the graded extension of
Poisson-sigma models also opens the door towards a similar treatment
of supergravity-like theories \cite{Ertl:2000si}. A subset of
those graded Poisson-sigma models (MFS in fig.\ \ref{fig:relations}), already identified
by the present authors as {}``genuine'' supergravities from their
algebra of Hamiltonian constraints \cite{Bergamin:2002ju}, is equivalent to
a second order superdilaton theory (MFDS). The latter, in a more
round-about way (path $\tilde{B}' \rightarrow D' \rightarrow \tilde{B}''$ in
fig.\ \ref{fig:relations}) is shown here to be \emph{identical} to the superfield dilaton theory
of ref.\ \cite{Park:1993sd} (SFDS in fig.\ \ref{fig:relations}), when
different auxiliary fields in the latter
are eliminated, when certain conformal transformations are made and
when the gravitino field is redefined appropriately. We also were
able to prove that during these procedures the corresponding supersymmetric
transformations are mapped exactly upon each other -- with a local
Lorentz transformation mixed in only. The chain of formulas which
provides the identification of superfield dilaton supergravity \cite{Park:1993sd} with
the corresponding class of supergravity theories derived from the
graded Poisson-sigma approach is compiled at the end of Section \ref{sec:torsiongrav}.
These formulas should be consulted together with a flow diagram of
fig.\ \ref{fig:relations}  and the corresponding list of actions.

Of course, once this identification is established, the general solution
of MFS, obtained after a conformal transformation (target space diffeomorphism
in gPSM parlance) from the solution published already in ref.\
\cite{Ertl:2000si}, represents the general solution of superfield supergravity
of ref.\ \cite{Park:1993sd} (cf.\ Section \ref{sec:solution}). Our only technical restriction
has been, that a further arbitrary function of the dilaton superfield $\Phi$ ($J(\Phi)$
in the action of ref.\ \cite{Park:1993sd}, resp.\ our \eqref{eq:ps1}) has been assumed
to be invertible so that the replacement $J(\Phi)\rightarrow\Phi$
is possible. This restriction is not serious as on the one hand all
physically interesting theories (spherically reduced gravity etc.)\
are of this form anyhow. On the other hand, this inversion is permitted
locally in function space, so that the only further complication can be that the general solution is
valid only in a certain patch in this case.

This general solution of superfield supergravity (after a suitable choice of
gauge-fixings among the bosonic symmetries) depends on a bosonic constant
Casimir function $C$ and on two bosonic functions
$\extd F$ and $\phi$, which may be interpreted as the coordinates of the world
sheet. For $C \neq 0$ the anti-commuting space is parametrized by two free
fermionic functions, at $C \equiv 0$ one of them is replaced by an anti-commuting
constant Casimir function and an anti-commuting $\extd f$.

As argued in ref.\ \cite{Strobl:1999zz} the
two fermionic gauge-degrees of freedom of $N=(1,1)$ supergravity could
be used locally to put those functions to zero (zero fermion gauge,
ZF gauge) so that the bosonic solution survives after all as the only
{}``non-trivial'' one, as long as no interactions with matter are considered. 
Indeed, by only demanding a non-singular determinant of the supermetric --
generalizing the non-singularity restriction of bosonic gravity --
this ZF gauge would be permitted, and, in this philosophy, pure supergravity
would become {}``trivial'', because the gravitino can be {}``gauged
away''. However, whether such a gauge is ``physically'' accessible depends on
the way, a ``measuring device'' (e.g.\ super-geodesic) is transformed in this
process. Only for a very restricted class of solutions an ``invariant'' device
can be constructed, namely when such a solution does not break
supersymmetry. In the generic case our solution, as well as the bosonic one of
ref.\ \cite{Park:1993sd}, show explicit supersymmetry breaking. The general
solution provided in the present work covers all
cases, but in the applications presented in our present paper -- for simplicity -- backgrounds in the
ZF gauge are considered, only.

The formulation of the massive super-point particle, moving along a {}``super-geodesic''
is straightforward in the superfield formalism, but so far -- for lack of a
general solution of the supergravity background -- has been very difficult
to work with in practice. Mapping the action of that particle into the
gPSM-based formalism by means of the identification
found in our present work, leads to a system of supergeodesic equations
(Section \ref{sec:sparticle}) which describes the motion in a background
solution. Interestingly enough,
even in the ZF gauge (with gravitino set to zero identically) the
bosonic geodesics (the {}``body'') are accompanied by an orbit in
the superpartner ({}``soul'') of the bosonic coordinates. This opens
an interesting field of detailed investigations of such systems where
body and soul may mutually influence each other in
the quest for a globally complete solution (described by a {}``super-Penrose diagram'').
We only present some simple examples here with massless (light-like)
movement and especially concentrate also on theories with Minkowski
ground state models in their bosonic sector. Such backgrounds become
flat for $C=0$. They include e.g.\ the Schwarzschild solution, but
also other models which are not asymptotically flat. 

Clearly the range of further studies, made possible by our present
work is very broad, because the (g)PSM technology is extremely powerful,
not only at the classical, but in particular also at the quantum level
\cite{Kummer:1992rt,Haider:1994cw,Schaller:1994pm,Schaller:1994np,Strobl:1994eu,Kummer:1997hy,Kummer:1998jj,Kummer:1998zs,Grumiller:2002nm}.
On the other hand, superspace techniques allow to couple matter fields in a
straightforward way to MDFS and NDMFS (cf.\ 
fig.\ \ref{fig:relations}). Therefore a path integral quantization of MFS
coupled to matter fields becomes manageable
\cite{Bergamin:2001}. Also questions like the one concerning a fermionic
counterpart of the (bosonic) virtual Black Hole \cite{Grumiller:2000ah,Grumiller:2002dm} can
be expected to find an answer - as well as a plethora of further research
directions, e.g. the inclusion of Yang-Mills fields, just to name
one of them \cite{Kummer:1995qv,Guralnik:2003we,Grumiller:2003ad}.
\subsection*{Acknowledgement}
The authors are grateful to D. Grumiller for
numerous stimulating discussions. They thank T. Strobl and P. van
Nieuwenhuizen for illuminating comments. The work has been supported by the
Swiss National Science Foundation (SNF) and the Austrian Science Foundation
(FWF) Project 14650-TPH and P-16030-N08.

\appendix
\section{Appendix}
\label{sec:appendix}
\subsection{Notations and Conventions}
\label{sec:notation}
These conventions are identical to
\cite{Ertl:2000si,Ertl:2001sj}, where additional explanations can be found.

\vspace{2ex}
\noindent\emph{Generic} indices (used in the context of gPSM's) are
  chosen from the middle of the alphabet:
  \begin{itemize}
  \item $I$, $J$, $K$, \ldots include both commuting and anti-commuting
  objects. Generalized commutation relations are written in the standard way
  \begin{equation}
    v^I w^J = (-1)^{IJ} w^J v^I\ ,
  \end{equation}
  where the indices in the exponent take values 0 (commuting object) or 1
  (anti-commuting object).
  \item $i$, $j$, $k$, \ldots are generic commuting indices
  \end{itemize}

\noindent To label \emph{holonomic} coordinates, letters from the middle of the
  alphabet are used:
  \begin{itemize}
  \item $M$, $N$, $L$, \ldots can be both, commuting and anti-commuting.
  \item $m$, $n$, $l$, \ldots are commuting.
  \item $\mu$, $\nu$, $\rho$, \ldots are anti-commuting.
  \end{itemize}

\noindent\emph{Anholonomic} coordinates are labeled by letters from the
  beginning of the alphabet:
  \begin{itemize}
  \item $A$, $B$, $C$, \ldots can be both, commuting and anti-commuting.
  \item $a$, $b$, $c$, \ldots are commuting.
  \item $\alpha$, $\beta$, $\gamma$, \ldots are anti-commuting.
  \end{itemize}

\noindent The index $\phi$ is used to indicate the dilaton component of
  the gPSM fields:
  \begin{align}
    X^\phi &= \phi & A_\phi &= \omega
  \end{align}

The summation convention is always $NW \rightarrow SE$, especially for a
fermion $\chi$: $\chi^2 = \chi^\alpha \chi_\alpha$. Our conventions are
arranged in such a way that almost every bosonic expression is transformed
trivially to the graded case when using this summation convention and
replacing commuting indices by general ones. This is possible together with
exterior derivatives acting \emph{from the right}, only. Thus the graded
Leibniz rule is given by
\begin{equation}
  \label{eq:leibniz}
  \mbox{d}\left( AB\right) =A\mbox{d}B+\left( -1\right) ^{B}(\mbox{d}A) B\ .
\end{equation}

In terms of anholonomic indices the metric and the symplectic $2 \times 2$
tensor are defined as
\begin{align}
  \eta_{ab} &= \eta^{ab} = \left( \begin{array}{cc} 1 & 0 \\ 0 & -1
  \end{array} \right)\ , &
  \epsilon_{ab} &= - \epsilon^{ab} = \left( \begin{array}{cc} 0 & 1 \\ -1 & 0
  \end{array} \right)\ , & \epsilon_{\alpha \beta} &= \epsilon^{\alpha \beta} = \left( \begin{array}{cc} 0 & 1 \\ -1 & 0
  \end{array} \right)\ .
\end{align}
The metric in terms of holonomic indices is obtained by $g_{mn} = e_n^b e_m^a
\eta_{ab}$ and for the determinant the standard expression $e = \det e_m^a =
\sqrt{- \det g_{mn}}$ is used. The volume form reads $\epsilon = \half{1}
\epsilon^{ab} e_b \wedge e_a$.

The $\gamma$-matrices are used in a chiral representation:
\begin{align}
\label{eq:gammadef}
  {{\gamma^0}_\alpha}^\beta &= \left( \begin{array}{cc} 0 & 1 \\ 1 & 0
  \end{array} \right) & {{\gamma^1}_\alpha}^\beta &= \left( \begin{array}{cc} 0 & 1 \\ -1 & 0
  \end{array} \right) & {{\gthree}_\alpha}^\beta &= {(\gamma^1
    \gamma^0)_\alpha}^\beta = \left( \begin{array}{cc} 1 & 0 \\ 0 & -1
  \end{array} \right)
\end{align}
The matrices $(\gamma^a)^{\alpha \beta} = \epsilon^{\alpha \delta}
{{\gamma^a}_\delta}^\beta$ and $(\gthree)^{\alpha \beta}$ are symmetric in $\{
\alpha, \beta \}$.
The most important relations among the $\gamma$-matrices are:
\begin{align}
  \gamma^a \gamma^b &= \eta^{ab} \mathbf{1} + \epsilon^{ab} \gthree &
  \gamma^a \gthree &= \gamma^b {\epsilon_b}^a
\end{align}

Covariant derivatives of anholonomic indices with respect to the geometric
variables $e_a = \extd x^m e_{am}$ and $\psi_\alpha = \extd x^m \psi_{\alpha m}$
include the two-dimensional spin-connection one form $\omega^{ab} = \omega
\epsilon^{ab}$. When acting on lower indices the explicit expressions read
($\half{1} \gthree$ is the generator of Lorentz transformations in spinor space):
\begin{align}
\label{eq:A8}
  (D e)_a &= \extd e_a + \omega {\epsilon_a}^b e_b & (D \psi)_\alpha &= \extd
  \psi_\alpha - \half{1} {{\omega \gthree}_\alpha}^\beta \psi_\beta
\end{align}

Finally light-cone components are introduced. As we work with spinors in a
chiral representation we can use
\begin{align}
\label{eq:Achi}
  \chi^\alpha &= ( \chi^+, \chi^-)\ , & \chi_\alpha &= \begin{pmatrix} \chi_+ \\
  \chi_- \end{pmatrix}\ .
\end{align}
For Majorana spinors upper and lower chiral components are related by $\chi^+
= \chi_-$, $ \chi^- = - \chi_+$, $\chi^2 = \chi^\alpha \chi_\alpha = 2 \chi_- \chi_+$. Vectors in light-cone coordinates are given by
\begin{align}
\label{eq:A10}
  v^{++} &= \frac{i}{\sqrt{2}} (v^0 + v^1)\ , & v^{--} &= \frac{-i}{\sqrt{2}}
  (v^0 - v^1)\ .
\end{align}
Derivatives with respect to these components are
written compactly as
\begin{equation}
\label{eq:Ader}
  \partial_K = \derfrac{X^K} = (\partial_\phi, \partial_{++}, \partial_{--},
  \partial_{+}, \partial_{-})\ .
\end{equation}
The additional factor $i$ in \eqref{eq:A10} permits a direct identification of the light-cone components with
the components of the spin-tensor $v^{\alpha \beta} = \frac{i}{\sqrt{2}} v^c \gamma_c^{\alpha
  \beta}$. This implies that $\eta_{++|--} = 1$
and $\epsilon_{--|++} = - \epsilon_{++|--} = 1$. The
$\gamma$-matrices in light-cone coordinates become
\begin{align}
\label{eq:gammalc}
  {(\gamma^{++})_\alpha}^\beta &= \sqrt{2} i \left( \begin{array}{cc} 0 & 1 \\ 0 & 0
  \end{array} \right)\ , & {(\gamma^{--})_\alpha}^\beta &= - \sqrt{2} i \left( \begin{array}{cc} 0 & 0 \\ 1 & 0
  \end{array} \right)\ .
\end{align}
\subsection{Superspace Integration}
\label{sec:integrals}
As for fermionic fields the square of the superspace variables is abbreviated
by $\theta^2 = \theta^\alpha \theta_\alpha$. Superspace is integrated out by
\begin{equation}
\label{eq:thetaint}
  \half{1} \intd{\diff{\theta^2}} \theta^2 = 1\ .
\end{equation}
Our covariant derivative with respect to anholonomic indices is:
\begin{equation}
  D_\alpha = \partial_\alpha + i (\gamma^a \theta)_\alpha \partial_a
\end{equation}
With these conventions and the particular choice of gauge used in this work
the components\footnote{Field components in a superfield are denoted by
  underlined symbols throughout.} of the super-zweibein read \cite{Ertl:2000si}
\begin{align}
\label{eq:szb1}
  {E_m}^a &= {e_m}^a + 2i (\theta \gamma^a \underline{\psi}_m) + \half{1} \theta^2 \underline{A}
  {e_m}^a\ ,\medsp
\label{eq:szb2}
  {E_m}^\alpha &= {\underline{\psi}_m}^\alpha - \half{1} \underline{\tilde{\omega}}_m (\theta \gthree)^\alpha
  + \half{i} \underline{A} (\theta \gamma_m)^\alpha - \half{1} \theta^2 \bigl( \frac{3}{2}
  \underline{A} {\underline{\psi}_m}^\alpha + i (\underline{\tilde{\sigma}} \gamma_m \gthree)^\alpha - \underline{A} (\underline{\zeta}
  \gamma_m)^\alpha \bigr)\ , \medsp
\label{eq:szb3}
  {E_\mu}^a &= i (\theta \gamma^a)_\mu\ , \medsp
\label{eq:szblast}
  {E_\mu}^\alpha &= {\delta_\mu}^\alpha \bigl(1 - \inv{4} \theta^2
  \underline{A} \bigr) \ ,
\end{align}
where $\underline{\tilde{\omega}}$ and $\underline{\tilde{\sigma}}$ are
defined as in \eqref{eq:tildeomega} and \eqref{eq:susysigma}, respectively,
however expressed in terms of underlined fields.
In superspace it is often useful to introduce the
Lorentz covariant decomposition of the gravitino field
\begin{align}
\label{eq:gravitinodecomp}
  \underline{\psi}^a_\alpha &= (\underline{\zeta} \gamma^a)_\alpha + \underline{\lambda}^a_\alpha\ , &
  \underline{\zeta}_\alpha &= \half{1} (\underline{\psi}^a \gamma_a)_\alpha\ , &
  \underline{\lambda}^a_\alpha &= \half{1} (\underline{\psi}^b \gamma^a \gamma_b)_\alpha\ .
\end{align}
Finally we provide the relevant superspace integrations in the general
dilaton model of ref.\ \cite{Park:1993sd}. According to eq.\ \eqref{eq:ps1} the
following integrations have to be performed
\begin{equation}
\label{eq:appexp1}
  \intd{\diff{^2 \theta}} E J(\Phi) S = e \biggl(\half{1} \underline{\tilde{R}} J + J'
  (\underline{\chi} \underline{\tilde{\sigma}}) + J' \underline{F}
  \underline{A} + \inv{8} J'' \underline{A} \underline{\chi}^2 \biggr)\ ,
\end{equation}
\begin{multline}
\label{eq:appexp2}
  \intd{\diff{^2 \theta}} E K(\Phi) D^\alpha \Phi D_\alpha \Phi =\medsp = e \biggl( 2
  K \bigl( \partial^m \underline{\phi} \partial_m \underline{\phi} - \frac{i}{4} \underline{\chi} \gamma^m
  \partial_m \underline{\chi} + \underline{F}^2 - (\underline{\psi}_n \gamma^m
  \gamma^n \gthree \underline{\chi}) \partial_m \underline{\phi} \bigl) \medsp
   + \inv{4} K \underline{\chi}^2 (\underline{\psi}_n \gamma^m \gamma^n \underline{\psi}_m)  +
  \inv{4} K' \underline{\chi}^2 \underline{F} \biggr)\ ,
\end{multline}
\begin{equation}
\label{eq:appexp3}
  \intd{\diff{^2 \theta}} E L(\Phi) = e \biggl( \bigl(\underline{A} + 2 \underline{\zeta}^2 +
  \underline{\lambda}^2 \bigr) L + L' \bigl( \underline{F} + i (\underline{\zeta}\gthree\underline{\chi})\bigl) + \inv{8} L''
  (\underline{\chi} \underline{\chi}) \biggr)\ .
\end{equation}
Evaluating \eqref{eq:ps1} with \eqref{eq:appexp1}-\eqref{eq:appexp3},
the variation of the action with
respect to the  auxiliary fields $\underline{A}$ and $\underline{F}$ yields the elimination
conditions for those fields:
\begin{align}
\label{eq:psauxfield1}
  \underline{F} &= - \inv{J'} \bigl( L + \inv{8} J'' \underline{\chi}^2 \bigr)\ , \medsp
  \label{eq:psauxfield2} \underline{A} &=
   4 \frac{KL}{(J')^2} - \frac{L'}{J'} + \bigl(\frac{K J''}{2(J')^2} - \inv{4}
   \frac{K'}{J'}\bigr) \underline{\chi}^2 \ . 
\end{align}

\providecommand{\href}[2]{#2}\begingroup\raggedright\endgroup

\end{document}